\documentclass[conference]{IEEEtran}

\usepackage{graphicx}
\usepackage{amsmath}
\usepackage{caption}
\usepackage{epstopdf}
\usepackage{float}
\usepackage{xcolor}
\usepackage{amsfonts}
\begin{document}
%
\title{Low PAPR Reference Signal Transceiver Design for $3$GPP $5$G NR Uplink}

\author{\IEEEauthorblockN{M.~Sibgath Ali Khan, Sai Dhiraj~Amuru, Kiran Kuchi}
\IEEEauthorblockN{Department of Electrical Engineering\\
Indian Institute of Technology - Hyderabad, India\\
Email: \{ee13p0003, asaidhiraj, kkuchi\}@iith.ac.in\\}	\vspace{-20pt}}
\maketitle

\begin{abstract}\label{sec:Abstract}
Low peak-to-average-power ratio (PAPR) transmissions significantly improve the cell coverage as they enable high power transmissions without saturating the power amplifier. A new modulation scheme, namely, $\pi/2$-BPSK was introduced in the Rel-$15$ $3$GPP $5$G NR specifications to support low PAPR transmissions using the DFT-spread-OFDM waveform in the uplink transmissions. To enable data demodulation using this modulation scheme, Zadoff-Chu sequences are used as reference signals. However, the PAPR of Zadoff-Chu sequences is higher when compared to the $\pi/2$-BPSK data. Therefore, even though the data transmissions have low PAPR, the high PAPR of the reference signal limits the cell coverage in the uplink of Rel-$15$ $3$GPP $5$G NR design. In this paper we propose a transceiver design which minimizes the PAPR of the reference signals to avoid the aforementioned issues. We show via simulations that the proposed architecture results in more than $2$ dB PAPR reduction when compared to the existing design. In addition, when multiple stream transmission is supported, we show that PAPR of the reference signal transmission remains the same for any stream (also referred to as baseband antenna port in $3$GPP terminology) when the proposed transceiver design is employed, which is not the case for the current $3$GPP $5$G NR design.
\end{abstract}


\IEEEpeerreviewmaketitle
\vspace{-10pt}
\section{Introduction}\label{sec:Introduction}
For a cellular network, uplink transmissions define the coverage area. This is because the transmission power in the uplink is limited to $23$ dBm at the user equipment (UE) owing to hardware limitations (such a battery size) and regulatory constraints as opposed to $43$ dBm at the base station in the downlink \cite{101}. This limited transmission power in the uplink must therefore be used carefully to enhance cell coverage without increasing the CAPEX/OPEX costs of deploying more cell sites. Therefore the uplink design of a cellular standard is crucial in enabling uplink transmissions at high powers without saturating the power amplifier, which otherwise results in unwanted non-linear distortions. 

To address the above issues and to enhance the cell coverage of the newly designed $3$GPP $5$G NR when compared to $4$G LTE, a new modulation scheme, namely, $\pi/2$-BPSK was introduced for the uplink data channel (physical uplink shared channel- PUSCH) and control channel (physical uplink control channel - PUCCH) transmission. This waveform, when combined with an appropriate spectrum shaping enables low peak-to-average-power (PAPR) ratio transmissions without compromising the error rate performance \cite{211}-\cite{kk1}. Specifically, the PAPR of this modulation scheme with DFT-spread-OFDM waveform and spectrum shaping is smaller than $2$ dB. Moreover, it is shown in \cite{kk1}, \cite{kk2} that the power amplifier can be driven to saturation (adjacent channel leakage ratio (ACLR) and error vector magnitude (EVM) will still be within the required specification limits) and yet the error rate performance of this modulation scheme is not compromised. Hence, this modulation scheme plays a crucial role in significantly enhancing the cell coverage for $3$GPP $5$G NR-based cellular networks. 
shaping vector is performed

The demodulation reference signals (DMRS) employed in Rel-$15$ for coherent demodulation of the PUSCH and PUCCH are generated using Zadoff-Chu (ZC) sequences or QPSK-based Computer Generated Sequences(CGS) as specified in Section $5$.$2$.$2$ in \cite{211} and Section $6$.$2$.$2$ in \cite{213}. The PAPR of these sequences is around $3.5$-$4$ dB when spectrum shaping is employed which is higher than that of the spectrum-shaped data transmissions \cite{QC}\hspace{-0.5pt}-\cite{kk4}. Therefore, even though the data transmissions have low PAPR and potentially allow for larger coverage, the DMRS design still limits the cell size due to its high PAPR in Rel-$15$ $3$GPP $5$G NR. Note that, the performance of PUSCH and PUCCH channels directly depend on the quality of the channel estimates obtained using these DMRS sequences. Hence, when the DMRS sequences are transmitted at lower power to avoid PA saturation, the coverage of PUSCH and PUCCH channels is automatically limited. For this reason, $3$GPP introduced a new study item in Rel-$16$ to design new reference signal sequences with lower PAPR \cite{SI}. The sequences in \cite{3g}-\cite{Qual} were agreed to be used as low-PAPR reference sequences. In this paper, we will use them as the reference signal sequences for the proposed reference signal transceiver design.

\begin{figure}[h]
	\centering
	\includegraphics[width=0.7\columnwidth,height=5cm]{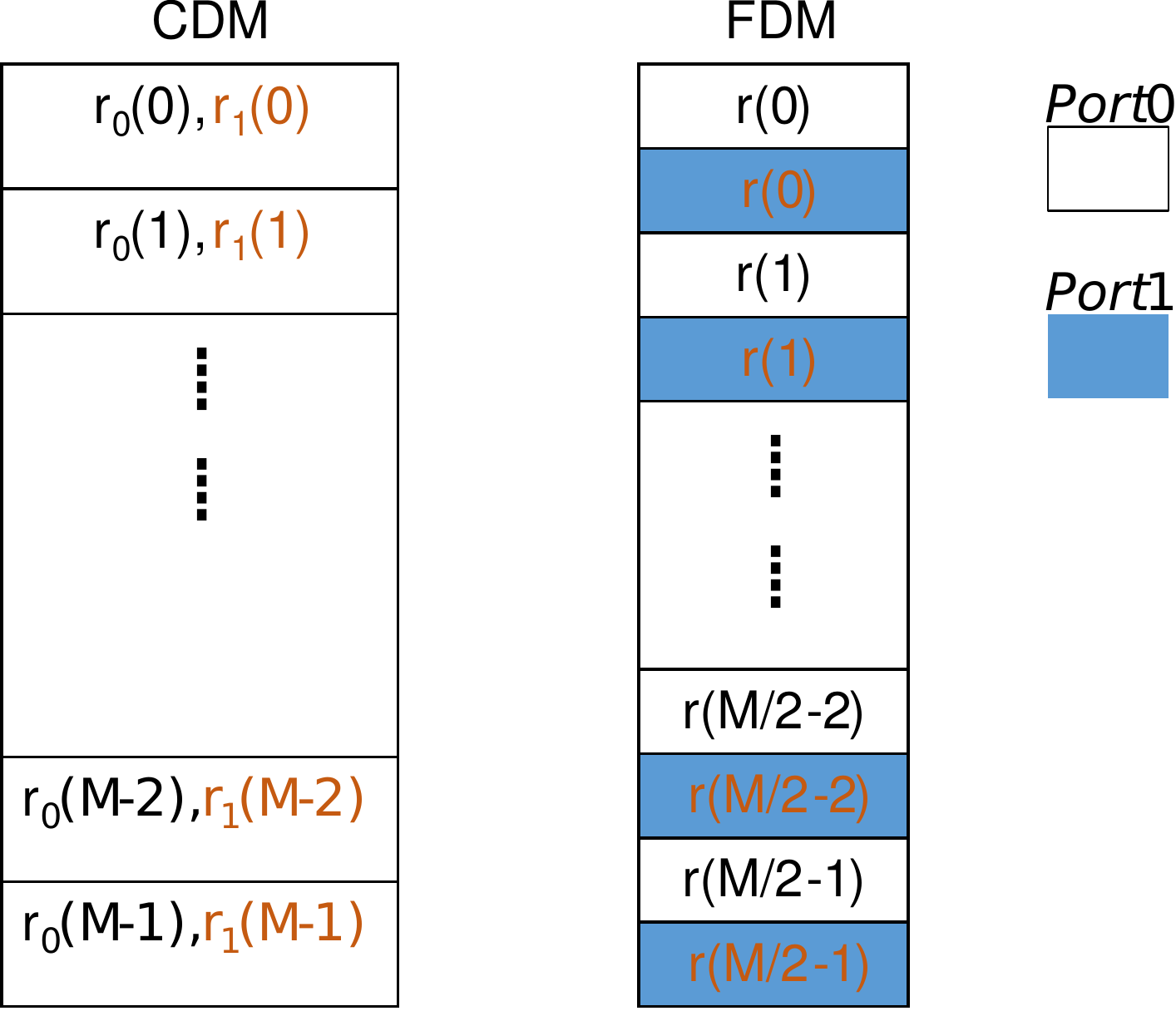}
	\caption{Port mapping for CDM, FDM method of reference signal multiplexing for MIMO stream transmissions in $3$GPP.}
	\label{fig:FDM_CDM}
	\vspace{-10pt}
\end{figure}

The Rel-$15$ specifications for $3$GPP $5$G NR also support multiple stream transmissions using DFT-spread-OFDM waveform. In other words, a single user can be scheduled to transmit multiple streams or multiple users can be configured simultaneously to transmit multiple streams depending on the channel conditions. In order to support these multiple-stream (also known as layers in $3$GPP terminology) MIMO transmission, multiple orthogonal DMRS sequences are necessary, one for each stream. This is achieved by introducing the concept of baseband antenna port where one single port is assigned for the demodulation of each stream/layer  \cite[Sec 6.3.1.3]{211}. Since the DMRS of each stream must be independently decoded for channel estimation of each stream, these DMRS sequences must be orthogonally separated to avoid any interference. In $3$GPP specifications, the orthogonality across the ports is achieved by frequency division multiplexing (FDM) or code division multiplexing (CDM). Distinct orthogonal DMRS sequences, each corresponding to an antenna port, share the same time-frequency resources in CDM method as shown in Fig.~\ref{fig:FDM_CDM} where $r_0$, $r_1$ are two distinct DMRS sequences corresponding to antenna port $0$ and antenna port $1$ respectively. In FDM method, the same sequence is employed for all the antenna ports but frequency multiplexed  as shown in Fig.~1b. It can be seen that in FDM the length of DMRS on each port will be $\frac{M}{P}$ rather than $M$, where $P$ indicates the number of antenna ports multiplexed in frequency domain. It is agreed in $3$GPP that Rel-16 NR \cite{3g} support only two layers via FDM and hence the length of DMRS on each port will be $\frac{M}{2}$ for a data allocation of length $M$ sub-carriers. We show in Section \ref{sec:ReceiverDesign} that this $\frac{M}{2}$-point reduction in DMRS length does not reduce the channel estimation quality and the $M$-length channel estimate vector corresponding to the $M$-length data allocation can be reconstructed perfectly. 

When multiple-stream transmissions are supported, the current $3$GPP Rel-$15$ specifications does not clearly mention the spectrum shaping implementation for the $\frac{\pi}{2}$-BPSK data and DMRS sequences. For instance, when multiple users each with one layer are configured to transmit simultaneously, a $\frac{M}{P}$ length DMRS sequence corresponding to each user's $M$ length data will be transmitted on one of the $P$ ports, in such case spectrum shaping has to align between data and DMRS  transmissons so that channel can be estimated correctly, which otherwise may result in imperfect receiver implementations (causing a loss of data exchanged). In addition to this, if proper design choices are not made, then it is also possible that the same DMRS sequence when mapped to two different baseband antenna ports (for example, as shown in Fig.~(1b)), it will behave differently with respect to (w.r.t) PAPR, auto and/or cross-correlation which eventually impact the channel estimation performance (immunity to inter-cell interference) and subsequently data demodulation. Therefore in this paper, we propose two transceiver architectures which generate low PAPR DMRS waveform and also results in identical channel estimation performance on all the baseband antenna ports. Specifically, we show the the sequences designed in \cite{3g}-\cite{Qual} to have low PAPR will have same error rate performance on any stream in the case of multiple-stream transmissions.
 

\textit{Notation:} The following notation is used in this paper. Upper case letters $\mathbf{X}$ denote matrices, bold lower case letters $\mathbf{x}$ denote vectors, non bold face letters represent scalars and $\mathbf{x}_t,\mathbf{y}_f$ indicates the time domain and frequency domain vectors $x$ amd $y$ respectively. $\mathbf{x}^T$ and $\mathbf{X}^{\dagger}$ represent the transpose and Hermitian operations on the vector $\mathbf{x}$ and matrix $\mathbf{X}$ respectively. We use the symbol $\mathbf{x}$ to denote the data symbols and $\mathbf{r}$ to denote reference signal symbols. 

\section{Transmitter Architecture for $\pi/2$-BPSK DATA and DMRS generation}\label{sec:SignalModel}
In this section, we present transmitter designs to generate low PAPR data, and DMRS waveforms.  We first describe the system model, including the design of the DFT-s-OFDM waveform as per the current 3GPP 5G NR specifications and then discuss the proposed transmitter designs.  

\subsection{DFT-s-OFDM Signal Model}\label{subsec:SignalModel}
In the current NR specifications \cite{211}, \cite{213}, Discrete Fourier transform-spread orthogonal frequency-division multiplexing (DFT-s-OFDM) \cite{Text1} is used for the uplink transmission, especially in coverage limited scenarios. This waveform is also referred to as single-carrier FDM waveform (SCFDM) in the literature. {In $3$GPP $5$G NR, QAM modulation symbols with modulation order $(4, 16, 64, 256)$ can be transmitted using the DFT-s-OFDM}. When compared to LTE, a new modulation scheme, namely, $\frac{\pi}{2}$-BPSK was introduced in $5$G NR. This is a special constellation-rotated BPSK modulation, such that even-numbered symbols are transmitted as in BPSK and the odd-numbered data symbols are phase rotated by $\frac{\pi}{2}$ as given below - 
        \begin{equation}\label{eq:pi/2 BPSK}
        	\hspace{-5pt}{x_p}(m)=\frac{(1+1i)}{\sqrt{2}}e^{i\hspace{0.5mm} (m\hspace{-5pt} \mod 2)\frac{\pi}{2}}x_t(m) ,\hspace{0.1cm}  m\in [0,\ldots,M-1],  
        \end{equation} 
        where $i=\sqrt{-1}$ and $M$ is the length of a BPSK sequence $x(m)$. Here the sub-script $p$ in $x_p(m)$ indicates a phase rotated sequence and the sub-script $t$ in $x_t(m)$ indicates a time-domain sequence. The $\frac{\pi}{2}$-phase rotation can be equivalently expressed in vector notation as given below 
        \begin{equation}\label{eq:pi/2 BPSK vector} 
        \mathbf{x}_p=\frac{1+i}{\sqrt{2}}\mathbf{P}\mathbf{x}_t 
        \end{equation}
        where $\mathbf{x}_t$ is a $M$ length BPSK vector, $P$ is $M\times M$ diagonal matrix with diagonal entries $p_{mm}=e^{i\hspace{0.5mm}(m \hspace{-10pt} \mod 2 \\)\frac{\pi}{2}}$.
        
\begin{figure}[h]
 	\centering
 		\includegraphics[width=0.8\columnwidth]{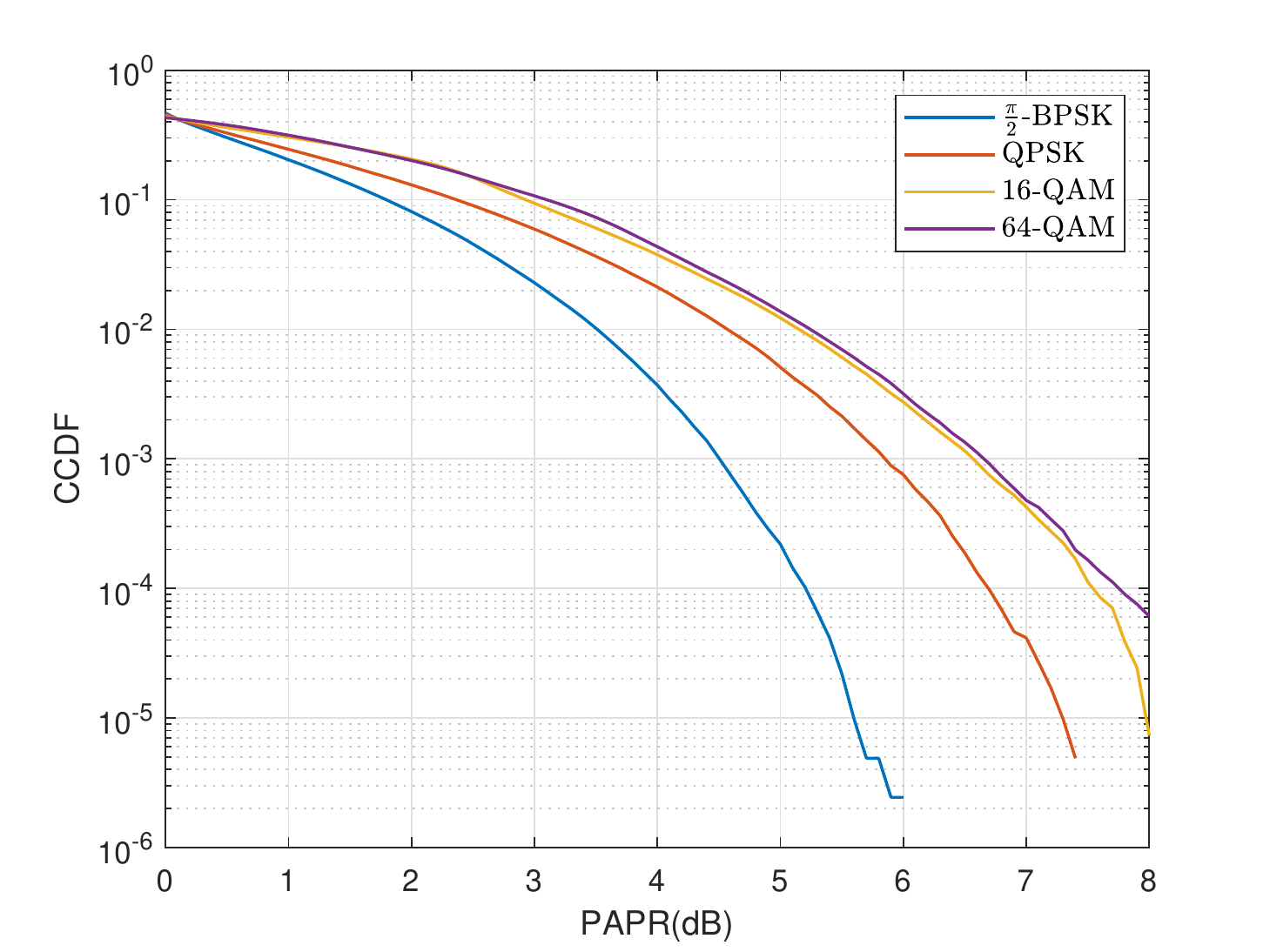}
 	 	\caption{PAPR of different modulation schemes using a DFT-s-OFDM waveform.}
 	\label{fig:PAPR_mods}
 	\vspace{-20pt}
\end{figure}

The $\frac{\pi}{2}$-BPSK modulation scheme when transmitted using DFT-s-OFDM has a low PAPR when compared to higher-order modulation schemes including QPSK as the zero-crossing transitions are avoided. The PAPR for various modulation schemes is shown in  Fig.~\ref{fig:PAPR_mods}, which clearly shows the low PAPR behavior of the $\frac{\pi}{2}$-BPSK modulation scheme. Note that, although the constellation is similar to QPSK, we can only transmit $1$-bit on one $\frac{\pi}{2}$-BPSK modulation symbol.
 %
 \begin{figure}[h]
 	\centering
	\includegraphics[width=0.8\columnwidth]{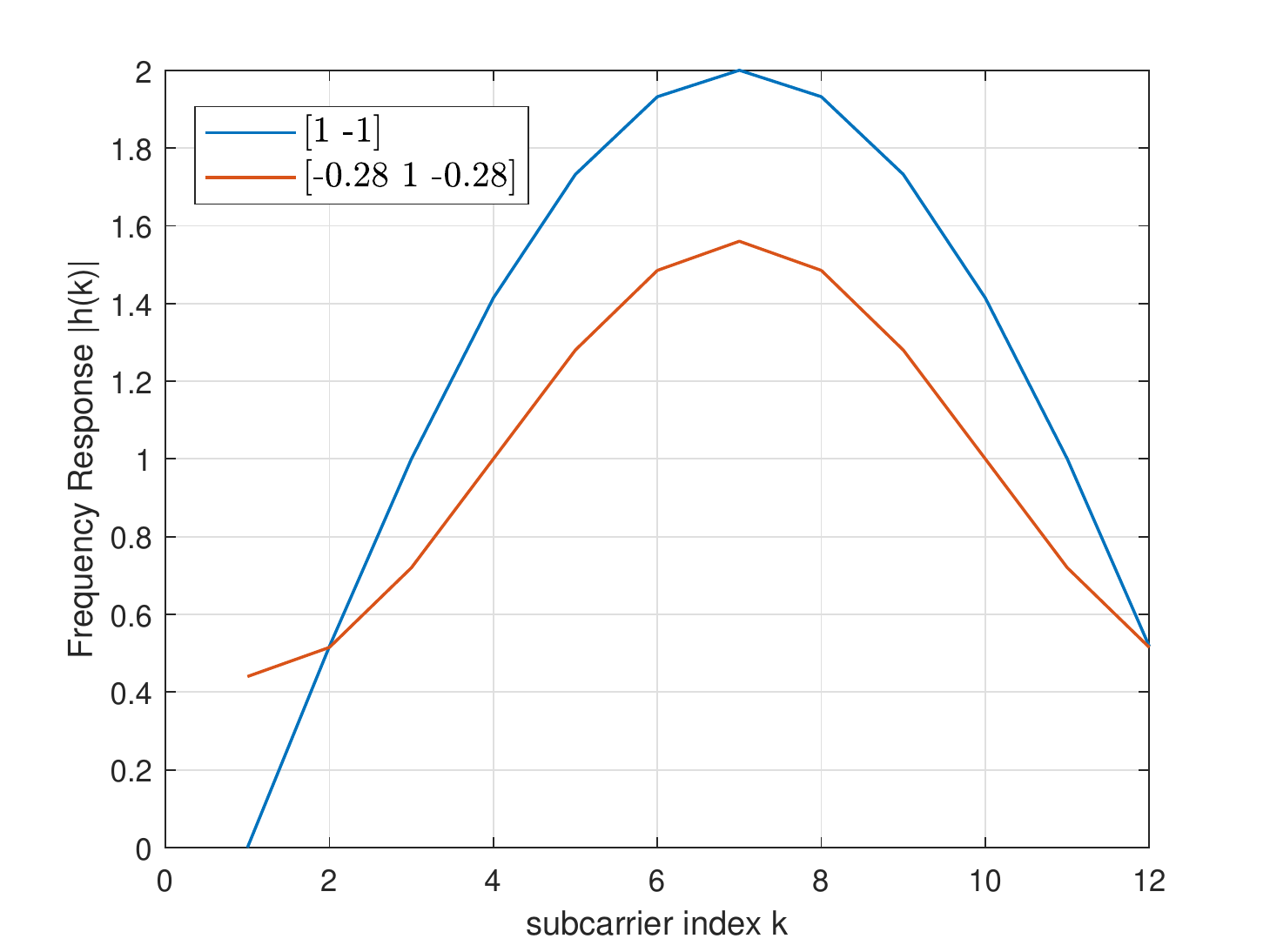}
	\caption{Frequency response of commonly used spectrum shaping filters with $2$-tap and $3$-tap impulse response.}
	\label{fig:spectrum_shaping}
	\vspace{-15pt}
\end{figure}

\subsection{Spectrum shaping}\label{sec:spectrum_shaping}
Spectrum shaping is a data-independent PAPR reduction technique which can be performed either in time domain or frequency domain \cite{kk1}, \cite{kk2}. In case of frequency-domain processing, spectrum shaping can be performed by means of a 
spectrum-shaping function $\mathbf{w}_f=\mathbf{D}_M\mathbf{w}_t$, where $\mathbf{w}_t$ is zero-padded time domain impulse response of the $L$-tap spectrum shaping filter i.e., $\mathbf{w}_t=[w(0),w(1),..w(L-1),\underbrace{0,\ldots,0}_{M-L}]^T$. Commonly used spectrum shaping filters with $2$ and $3$-tap impulse response are shown in Fig.~\ref{fig:spectrum_shaping}. 

\textbf{Remark on the length of the spectrum shaping filter:} \textit{In a recent study \cite{Korea}, a joint optimization of the rotation angle (other than $\frac{\pi}{2}$) and the spectrum shaping function is considered for further optimization of the PAPR of the BPSK-based DFT-s-OFDM waveforms beyond what is achieved using the filters shown in Fig.~\ref{fig:spectrum_shaping}. The spectrum shaping filter obtained via optimization in \cite{Korea} is of the length ranging between $8$-$24$. {To estimate the channel  at the receiver, in \cite{Korea} it is assumed that the spectrum shaping filter is perfectly known at the receiver and then the impulse response of the wireless channel is estimated for data demodulation. This violates the $3$GPP design} wherein it is clearly mentioned that the spectrum shaping filter is implementation-specific \cite{101} and therefore this filter is unknown at the receiver. In such cases, the receiver will have to estimate the joint impulse repsonse of the spectrum shaping filter and the wireless channel (will be explained in detail in Section~\ref{sec:chan_est_length_port}). Note that, a worst case wireless channel impulse response will be of length $\leq 3$ for an allocation of size $12$ subcarriers (i.e., $1$ resource block in $3$GPP terminology) as per $3$GPP channel models \cite{901}. Now, if the spectrum shaping filter is unknown at the receiver, we will need a minimum of $11$-$27$ samples to estimate the joint impulse response as per the design in \cite{Korea} which forces the data allocation to be a minimum of $2$-$4$ resource blocks (RB). Again this is contradicting the $3$GPP design where the minimum allocation size is $1$ RB. Hence, the length of the spectrum shaping filter has to be less than or equal to $3$ \cite{101} assuming two CDM groups with $6$ DMRS samples per CDM in a RB. Therefore, in this paper we restrict our analysis and simulations to filters with length $\leq 3$.} 
\begin{figure*}
	
	\centering
	\includegraphics[width=0.8\textwidth]{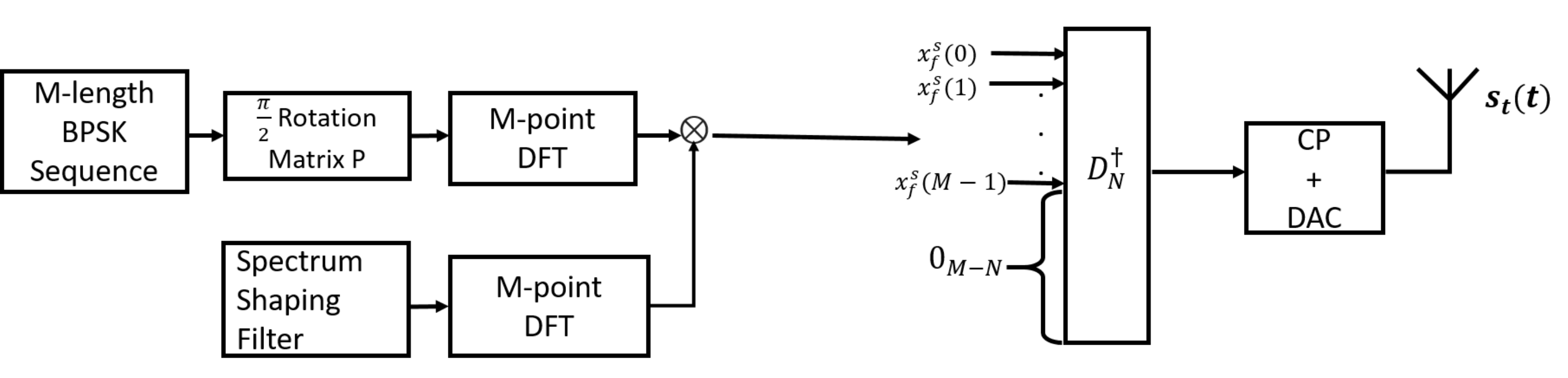}
	\caption{Transmitter architecture for data waveform generation using method-$1$.}
	\label{fig:Data_M1}
	\vspace{-12pt}
\end{figure*}
\subsection{DMRS Signal Structure}
As discussed in Section~\ref{sec:Introduction}, multiple DMRS sequences are transmitted on frequency division multiplexed antenna ports \cite{211}, \cite{213} to support MIMO transmissions. It should be noted that if spectrum shaping is performed on data symbols, identical spectrum shaping should also be performed on DMRS sequences to facilitate proper channel estimation and thereby equalization. However, if this spectrum shaping is not done in the right manner, will alter the properties of the DMRS waveform depending on the antenna port on which DMRS sequence is transmitted, which subsequently may result in non-identical channel estimation (and thereby equalization and demodulation) performance across the antenna ports which is not desirable.

Hence the DMRS transmitter design, besides minimizing the PAPR of the waveform should also ensure that the characteristics of the waveform (like auto-correlation and cross-correlation) are similar for spectrum-shaped DMRS sequences across all the antenna ports. In this paper, we propose two transmitter designs such that the PAPR of DMRS waveform is low and also the characteristics of the waveform are uniform across all the baseband antenna ports.

{In the current $3$GPP specifications \cite{3g}, $2$ MIMO streams are supported when $\frac{\pi}{2}$-BPSK modulation scheme is used. To support two MIMO streams,} two FDM DMRS ports are most commonly used as opposed to CDM (wherein the code orthogonality may be impacted in heavy delay spread channels). For the case of CDM, the DMRS sequences are mapped on the same antenna port and hence both DMRS ports are identical in terms of sequence generation, mapping and have same PAPR. The FDM case presents a challenging problem that needs to be addressed as will be discussed below. For FDM, a $M$-length data sequence on a given antenna port is associated with a corresponding $\frac{M}{2}$-length DMRS sequence. 

\subsection{Transmission Method - $1$}
In this section, we present data and DMRS transmission method-$1$ wherein the spectrum shaping is performed in the frequency domain. 
\subsubsection{\textbf{Data waveform design method-$1$}}
Let $ \mathbf{x}_t $ denote a $M \times 1$ vector of $\frac{\pi}{2}$-BPSK modulated data symbols generated as per  \eqref{eq:pi/2 BPSK}. For transmission via DFT-s-OFDM, the $\frac{\pi}{2}$-BPSK data symbols are first DFT-precoded as
 \begin{equation}\label{eq:dft equation}
  x_f(k)  =\sum_{m=0}^{M-1}x_p(m)e^{\frac{-i\hspace{0.5mm}2\pi km}{M}}.
 \end{equation}
The subscript $f$ in $x_f(k)$ indicates a frequency domain sequence. The DFT precoding shown in \eqref{eq:dft equation} can be equivalently represented in vector notation form as -
\begin{equation}\label{eq:dft-operation}
\mathbf{x}_f=\mathbf{D}_M\mathbf{x}_t,
\end{equation}
where $\mathbf{D}_M$ is a $M\times M$ DFT matrix given by 
\begin{equation*}
\mathbf{D}_M(k,m)=e^{\frac{-i\hspace{0.5mm}2\pi km}{M}}, 0 \leq k, m \leq M-1 
\end{equation*}

The spectrum shaping is performed on the DFT-precoded data vector as ${\mathbf{x}_f^s}=\mathbf{w}_f\mathbf{x}_f$, where ${\mathbf{x}_f^s}$ indicates the spectrum shaped frequency domain sequence $\mathbf{x}_f$. The spectrum-shaped data vector ${\mathbf{x}_f^s}$ is then mapped to a set of sub-carriers in frequency domain via a ${N\times M}$ mapping matrix $\mathbf{M}_f$ where $M\leq N$. The mapping matrix $\mathbf{M}_f$ is designed such that there are $M$ $1$'s in the matrix and $(N-1)M$ $0$'s, with the following constraints 
\begin{itemize}
\item There is a single location in each row that has $1$
\item No two rows can have $1$ in the same location
\item The total number of rows with $1$ is $M$
\end{itemize} 

The  mapping matrix $\mathbf {M}_f$ can be constructed such that it allocates $M$ sub-carriers in a localized or interleaved  manner. Finally, the output of this mapping operation is converted to $N\times 1$ time domain signal $\mathbf{s}_t$ as
\begin{equation*}
   \mathbf{s}_t=\mathbf{D}_N^{\dagger}\mathbf{M}_f {\mathbf{x}_f^s},
\end{equation*}
where $\mathbf{D}_N^{\dagger}$ is an inverse DFT matrix and $N$ is the total number sub-carriers corresponding to system bandwidth. 	
An appropriate length cyclic prefix is added to $\mathbf{s}_t$ to generate $\mathbf{s}_t(t)$ as given in equation (5.3.1) in 3GPP spec \cite{211}. This transmitter architecture for data waveform generation is shown in Fig.~\ref{fig:Data_M1}. 
     
    \begin{figure}
	
	\centering
	\includegraphics[width=0.8\columnwidth]{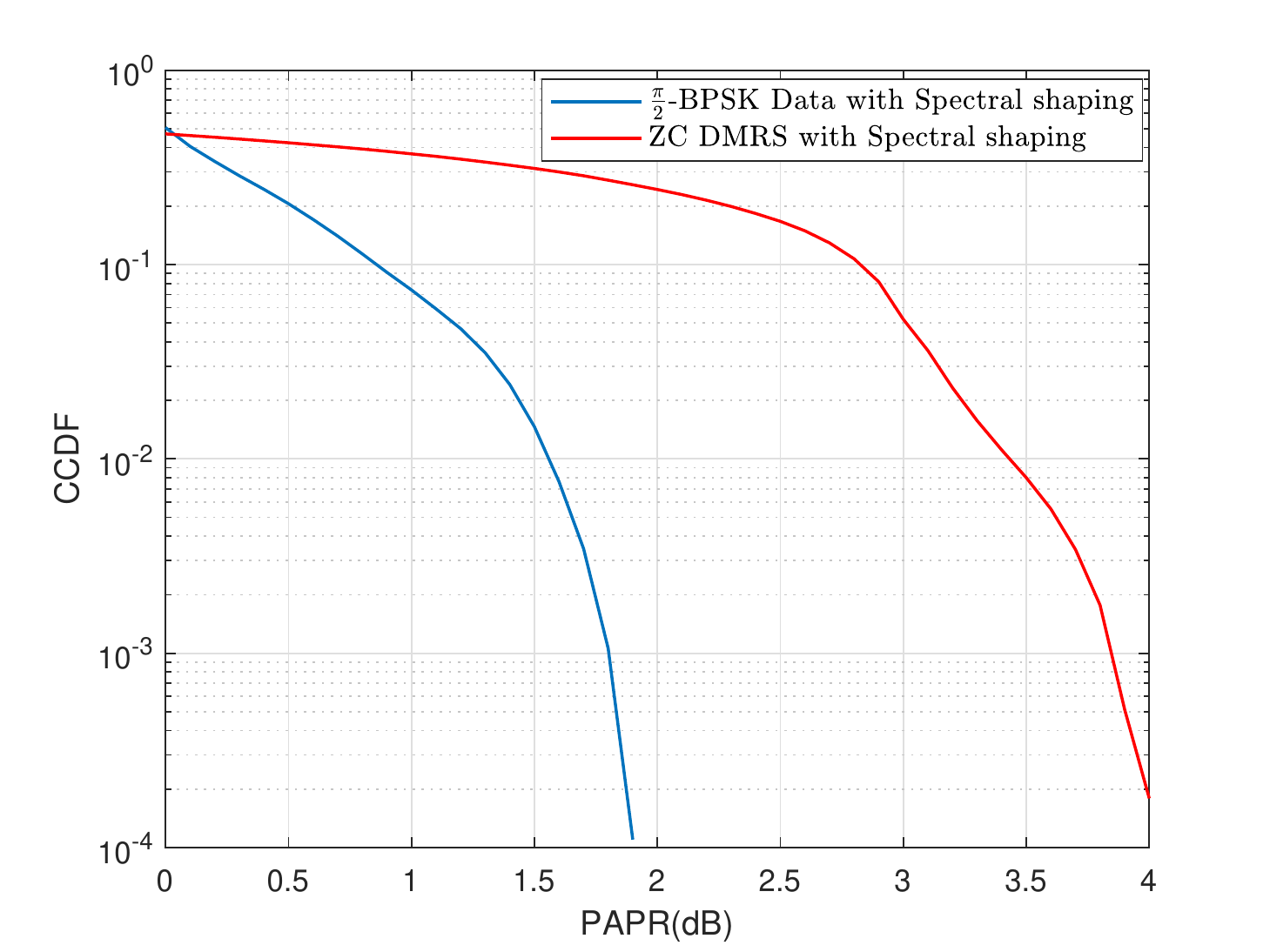}
	\caption{PAPR comparison between spectrum shaped ZC sequence and spectrum shaped $\frac{\pi}{2}$-BPSK data.}
	\label{fig:PAPR}
	\vspace{-12pt}
\end{figure}

\begin{figure*}
	\vspace{-10pt}
	\centering
	\includegraphics[width=0.8\textwidth]{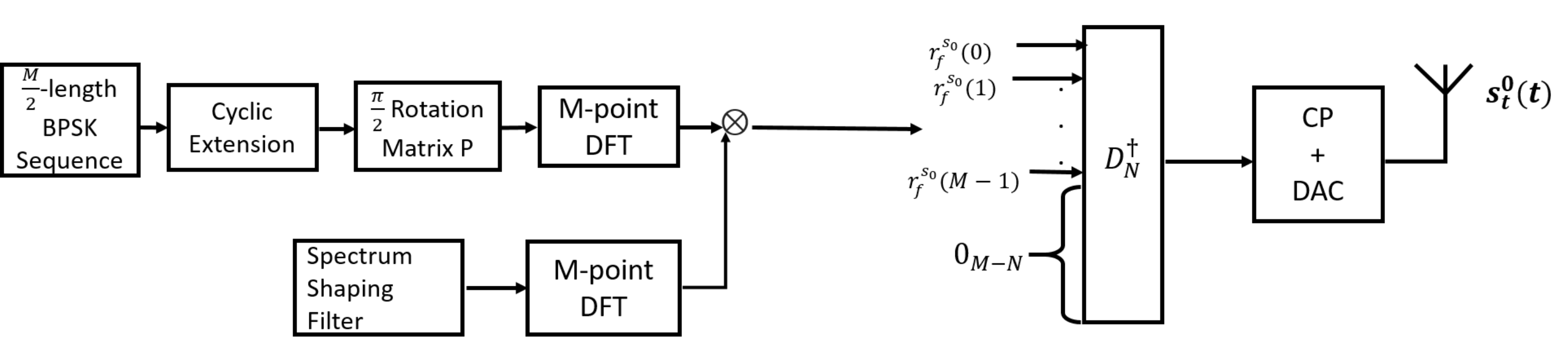}
	\caption{Transmitter architecture for \textit port-$0$ DMRS waveform generation using method-$1$. }
	\label{fig:DMRS_M1}
	\vspace{-10pt}
\end{figure*}

\begin{figure*}
	\vspace{-10pt}
	\centering
	\includegraphics[width=0.8\textwidth]{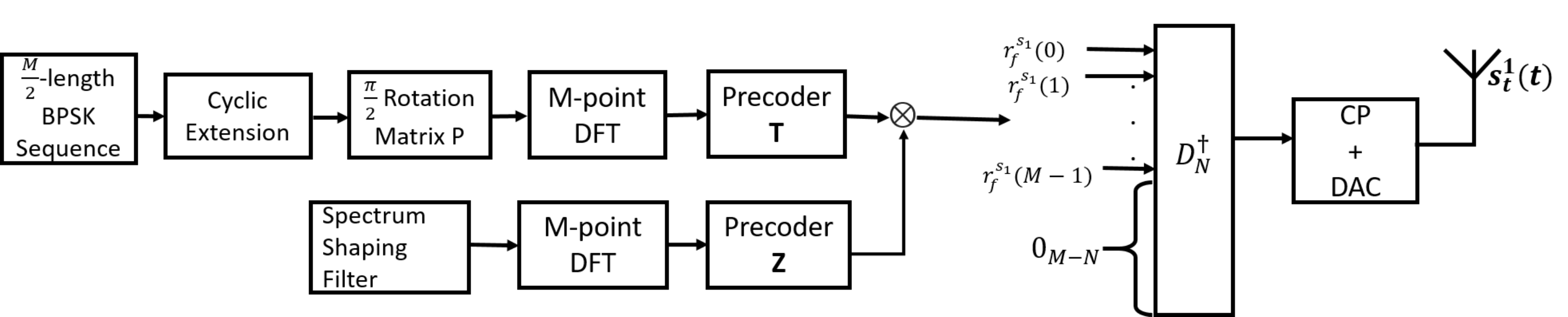}
	\caption{ Transmitter architecture for \textit port-$1$ DMRS waveform generation using method-$1$. }
	\label{fig:DMRS_M1_p1}
\end{figure*}

  \subsubsection{\textbf{DMRS waveform design method-$1$}}
 The CCDF of PAPR of a DFT-s-OFDM waveform with spectrum-shaped $\frac{\pi}{2}$-BPSK data symbols and the commonly used Zadoff-Chu based DMRS  sequences \cite [Section 5.2.2]{211}, \cite [Section 5.5]{213} is shown in Fig.~\ref{fig:PAPR}. It can be seen that PAPR of  $\frac{\pi}{2}$-BPSK is lower than that of the ZC sequences by over $2$dB. The high PAPR of ZC-based DMRS sequences will therefore limit the cell coverage as it is currently the case in Release $15$ $3$GPP $5$G NR. Hence there is a need for designing new reference signal sequences (DMRS) such that the PAPR of DMRS is similar to or lower than the data waveform. For this reason, 3GPP designed new DMRS sequences with low PAPR in \cite{SI}-\cite{Qual}. We will next describe how to use these sequences and design a transceiver to maintain the low PAPR for DMRS transmissions. As mentioned earlier, we assume $2$ MIMO streams are supported and the DMRS are multiplexed in an FDM manner for these streams. Hence, we assume $\frac{M}{2}$-length DMRS sequences will be transmitted for an $M$-length data allocation.

In this architecture the transmitter design is such that a given time domain DMRS signal $\mathbf{r}_t$ will result in an identical frequency domain signal $\mathbf{{r}}_f$  for any of the antenna ports. This subsequently results in similar auto and cross-correlation properties and hence produces an identical channel estimation performance at receiver. The system model of the architecture is shown in Figs.~\ref{fig:DMRS_M1}, \ref{fig:DMRS_M1_p1} and the summary is tabulated in Table \ref{tab:Method1} shown on the next page.
  
\textbf{DMRS waveform generation for Port 0}:    
  Let $\mathbf{r}_t$ be a pre-determined $\frac{M}{2}$-length DMRS sequence with BPSK modulated symbols chosen as per the designs in \cite{SI}-\cite{Qual}. This will be cyclically extended to result a $M$ length vector $\mathbf{\tilde{r}}_t (n)$ as follows 
	\begin{equation}
	\mathbf{\tilde{r}}_t (n)=\mathbf{r}_t\left(n\hspace{-10pt} \mod\frac{M}{2}\right), n=0,1,\ldots,M-1.
	\end{equation}
	
Using $\mathbf{P}$ defined in \eqref{eq:pi/2 BPSK vector}, a $\frac{\pi}{2}$-phase rotation is applied on $\mathbf {\tilde{r}}_t $ to give $\mathbf{\tilde{r}}_t^p=\mathbf{P}\mathbf{\tilde{r}}_t$. The resultant $\frac{\pi}{2}$-BPSK signal is DFT precoded as $\mathbf {r}_f^{p_0}=\mathbf{D}_M \mathbf{\tilde{r}}_t^p$. The resulting DFT-output will be a comb-like structure with non-zero entries only at odd locations which is equivalent to \textit{port}-$0$ mapping shown in Fig.\ref{fig:FDM_CDM} (and hence the notation $\mathbf {r}_f^{p_0}$). The DFT-precoded DMRS symbols are now spectrum-shaped using $\mathbf{w}_f$ defined in Section~\ref{sec:spectrum_shaping} to give the spectrum-shaped \textit{port}-$0$ DMRS as
 \begin{equation}
  {\mathbf{r}_f^{s_0}}=\mathbf{w}_f\mathbf{r}_f^{p_0}
  \end{equation}
  
   \textbf{DMRS waveform generation for Port 1}: 
   As per 3GPP specifications, in FDM-based multiplexing of multiple antenna ports, the DMRS sequence should be identical on both the ports i.e., the input BPSK sequence $\mathbf{r}_t$ and the resulting $\frac{\pi}{2}$-BPSK sequence $\mathbf{\tilde{r}}_t^p$ has to be same for both \textit{port}-0 and \textit{port}-1. However, different from \textit{port}-$0$, to generate the spectrum-shaped frequency domain-DMRS sequence on \textit{port}-$1$, the following additional steps need to be performed -
\begin{itemize}
\item a precoder $\mathbf T$ is applied on $\mathbf{\tilde{r}}_t^p$, where $\mathbf{T}$ is a $M\times M$-diagonal matrix with diagonal entries $T_{mm}=e^{i2\pi m/M }$ followed by DFT precoding as shown below
\begin{equation*}
  \mathbf {r}_f^{p_1}=\mathbf  {D}_M \mathbf T \mathbf{\tilde{r}}_t^p.
\end{equation*}
This $\mathbf {r}_f^{p_1}$ is a comb-like structure with non-zero entries only at even sub-carriers equivalent to \textit{port}-1 mapping as given in Fig.~\ref{fig:FDM_CDM}.
\begin{table*}[t]
	\centering
	{
		\caption{Summary of Method-$1$ based DMRS waveform generation}
		\label{tab:Method1}
		\renewcommand{\arraystretch}{1}
		{\fontsize{9}{9}\selectfont
			\begin{tabular}{|c|c|c|c|} \hline 
				Port &Time Domain DMRS &Spectrum shaping Filter &Freq Domain DMRS\\ 
				
				\hline \hline
				$0$ & 	$\mathbf{\tilde{r}}_t^p (n)=\mathbf{P}\mathbf{r}_t\left(n\hspace{-5pt} \mod\frac{M}{2}\right)$ &$\mathbf{w}_f=\mathbf{D}_M\mathbf{w}_t$ & $\mathbf{D}_M \mathbf{\tilde{r}}_t^p\mathbf{w}_f$   \\
				\hline
				$1$ & 	$\mathbf{\tilde{r}}_t^p (n)=\mathbf{T}\mathbf{P}\mathbf{r}_t\left(n\hspace{-5pt} \mod\frac{M}{2}\right)$&$\mathbf{w}_f=\mathbf Z\mathbf{D}_M\mathbf{w}_t$& $\mathbf{D}_M \mathbf{\tilde{r}}_t^p\mathbf{w}_f$  \\
				\hline
			\end{tabular}
		}
	}
\end{table*}
\item Spectrum shaping of $\mathbf {r}_f^{p_1}$ is done as follows - 
  \begin{equation*}
  \mathbf {r}_f^{s_1}=\mathbf Z \mathbf w_f \mathbf {r}_f^{p_1}
  \end{equation*}
  where, $\mathbf Z$ is a square-circulant matrix of size $M\times M$ whose $1$st row entries are $[\underbrace{0,0,...,0}_{M-1}, 1]$. 
\end{itemize}
  
\begin{figure}
 	\centering
	\includegraphics[width=0.8\columnwidth]{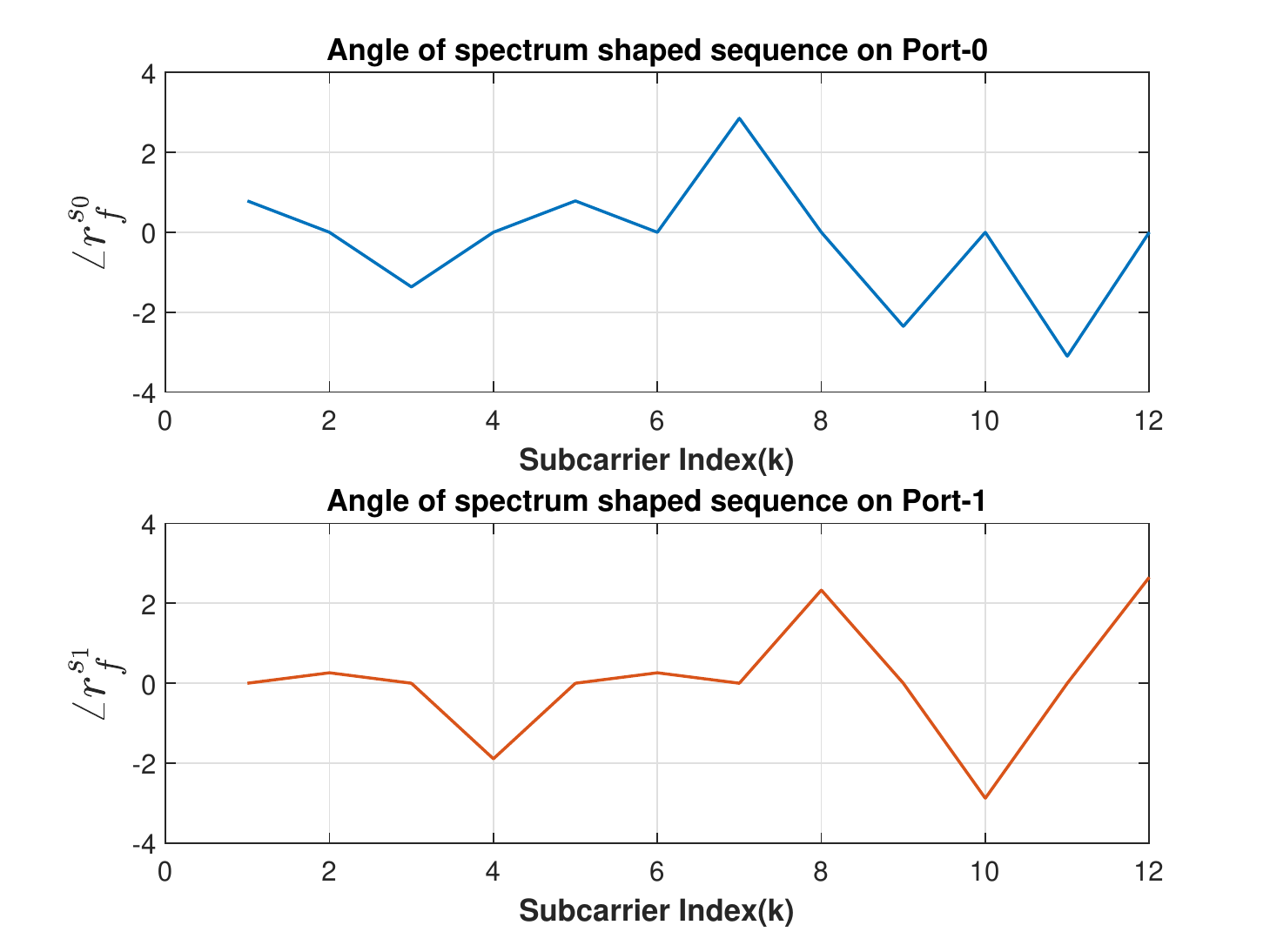}
	\vspace*{-5pt}
	\caption{Angle of $\mathbf {r}_f^{s_0}$, $\mathbf {r}_f^{s_1}$, i.e., the spectrum shaping filter outputs on \textit{port}-0 and \textit{port}-1 in the absence of precoder $\mathbf Z$. }
		\vspace{-10pt}
	\label{fig:Angle_P0_P1}
\end{figure}

{\textbf{Note:} Only when this precoder $\mathbf Z$ is applied on DMRS of port-$1$, the effect of spectrum shaping on data (shown in Fig.~\ref{fig:Data_M1}) and the DMRS on \textit{ports}-$0$, $1$ (shown in Figs.~\ref{fig:DMRS_M1}, \ref{fig:DMRS_M1_p1}) will be identical and data can be demodulated. In the absence of the precoder $\mathbf Z$, the non-zero entries of the spectrum shaped outputs $\mathbf {r}_f^{s_0}$, $\mathbf {r}_f^{s_1}$ will not be identical as shown in Fig.~\ref{fig:Angle_P0_P1}. This results in non-identical PAPR and channel estimation performance on \textit{port}-$0$ and \textit{port}-$1$, which is not acceptable in any MIMO system.}

Using the proposed architecture, it can be shown that the output of the spectrum shaping filter is identical for both the ports i.e.,
   \begin{align}\label{eq:M1 Equivalence}
   	 \mathbf{r}_f^{s_0}(2k) &= \mathbf{r}_f^{s_1}(2k+1) \nonumber \\
		&= \mathbf {r}_f^{p_0}(2k)\mathbf{w}_f(2k).
   \end{align}
   where $\mathbf {r}_f(k)$ is the M-point DFT of $\frac{\pi}{2}$-BPSK signal $\mathbf{\tilde{r}}_t^p$.
	Therefore, the same reference signal is transmitted on each baseband antenna port as per the $3$GPP $5$G NR specifications. {We further show in Section \ref{sec:ReceiverDesign} that the channel impulse response estimated on both the ports will be identical}.
	
 The spectrum-shaped DMRS vectors $ \mathbf {r}_f^{s_0}$, $\mathbf {r}_f^{s_1}$ are  mapped to a set of sub-carriers in frequency domain as discussed in Section~\ref{sec:spectrum_shaping}. The resulting output is converted to time domain via inverse-DFT operation similar to the method employed for data transmission as shown below -
   \begin{align}
  \mathbf s_t^0=\mathbf{D}_N^\dagger\mathbf{M}_f\mathbf {r}_f^{s_0}.\\
  \mathbf s_t^1=\mathbf{D}_N^\dagger\mathbf{M}_f\mathbf {r}_f^{s_1}.
  \end{align}

Using the above, the overall time-domain baseband signals $\mathbf{s}_t^0(t)$, $\mathbf{s}_t^1(t)$ with an appropriate cyclic prefix is generated as given by equation (5.3.1) in $3$GPP spec \cite{211}.

        
  \subsection{Transmission Method - $2$}
In the method-$1$ based transmitter design, the $\frac{\pi}{2}$-BPSK data and DMRS sequences are spectrum shaped in frequency domain. Further, the DFT-precoded DMRS sequences corresponding to each antenna port are generated and spectrum-shaped independently. In method-$2$ based design, we propose a low complexity design where spectrum shaping is performed in time-domain for both data and DMRS sequences via circular convolution operation. Specifically, a single DMRS sequence is spectrum-shaped in time-domain and mapped to both the antenna ports. The architecture for this transmitter design for the data and DMRS is shown in Figs.~\ref{fig:M2_Data} and \ref{fig:M2_DMRS} respectively. 
  
  \subsubsection{\textbf{Data waveform design method-$2$}}
{   
  \begin{figure*}
  	\centering
 	\includegraphics[width=0.8\textwidth]{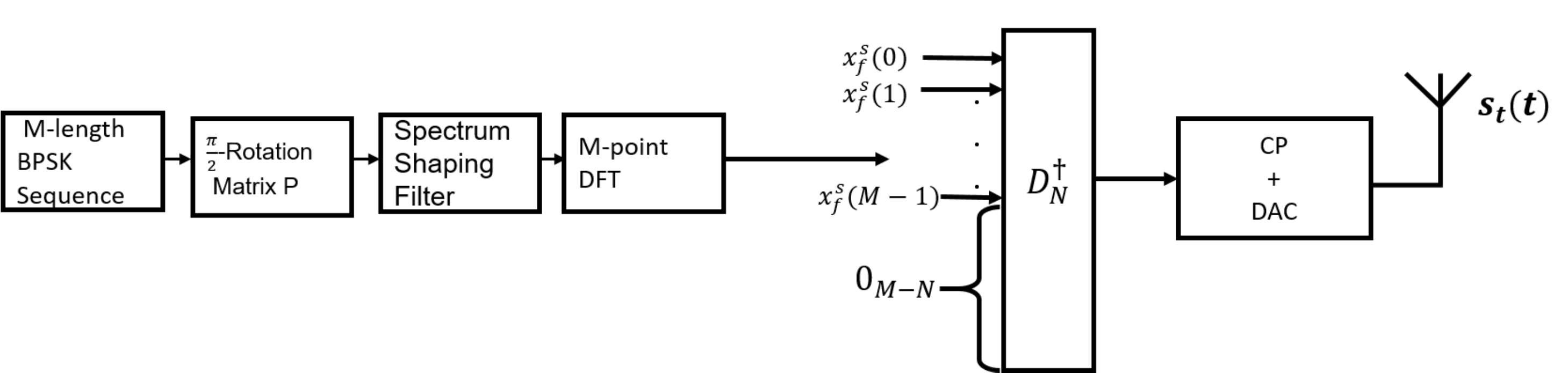}
 	\vspace*{-5pt}
 			\caption{Transmitter architecture for data waveform generation using method-$2$.}
 			 	\vspace{-5pt}
 		\label{fig:M2_Data}
 	\end{figure*}
 
  \begin{figure*}
  	\centering
 	\includegraphics[width=0.8\textwidth]{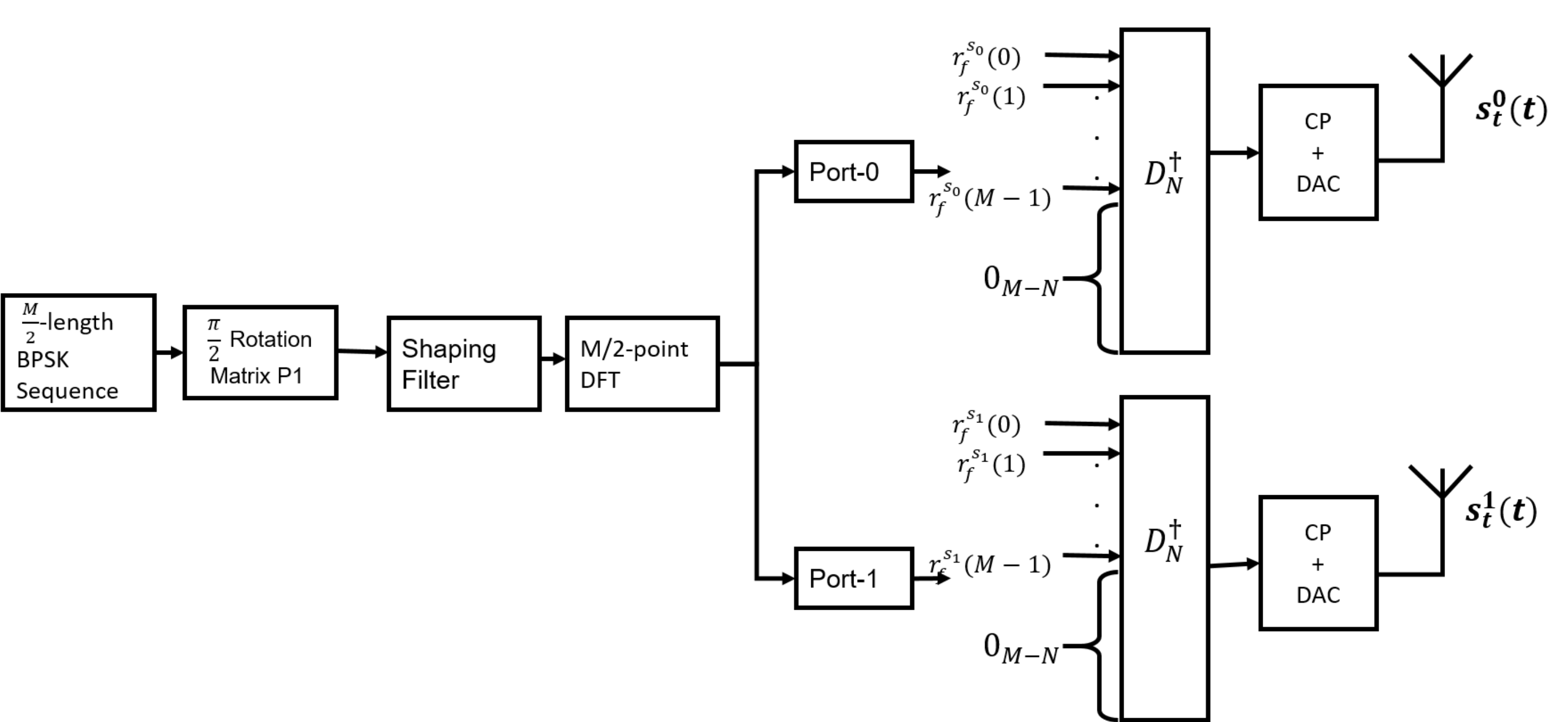}
 	\caption{ Transmitter architecture for DMRS waveform generation for \textit{port}-0 and \textit{port}-1 using method-$2$.}
 	 	\vspace{-10pt}
 	\label{fig:M2_DMRS}
 \end{figure*}
 
Let $\mathbf{x}_t$ be the $M$-length data vector to be transmitted from the UE to base station that  undergoes a $\frac{\pi}{2}$-phase rotation through a $M\times M$ diagonal matrix $P$. Here, $P$ is the same matrix used in method-$1$. This results in a $M$-length data vector $\mathbf x_t^p= \mathbf P \mathbf x_t$ with $\frac{\pi}{2}$-BPSK symbols. Note that in this method, the spectrum shaping of $\frac{\pi}{2}$-BPSK data,  is performed in time domain through a circular-convolution procedure with zero-padded $\mathbf{w}_t$ to produce a spectrum-shaped data as,
\begin{align}
\mathbf{x}_t^s(n)&=\sum_{m=0}^{M-1}\mathbf x_t^p(n)\mathbf{w}_t\left((m+n)\hspace{-10pt}\mod M\right),\nonumber \\
&\hspace{50pt} m, n \in \left[0,M-1\right]
\end{align}
The spectrum-shaped data  sequence is DFT precoded by means of $M$-point  as $\mathbf{x}_f^s =\mathbf{D}_M\mathbf{x}_t^s$.  The DFT precoded spectrum-shaped data vector is mapped to a set of sub-carriers in frequency domain via a mapping matrix $\mathbf{M}_f$ (described in Sec~\ref{sec:spectrum_shaping}). Finally, this mapped sequence is converted to time domain via inverse-DFT operation as
\begin{align}
\mathbf{s}_t =\mathbf{D}_N^{\dagger}\mathbf{M}_f\mathbf{x}_f^{s} \nonumber 
\end{align}
Using the above, the overall time-domain baseband signals $\mathbf{s}_t(t)$ with appropriate length cyclic prefix are generated as per equation (5.3.1) in 3GPP spec \cite{211}.
  \subsubsection{\textbf{DMRS waveform design method-$2$}}
 Let $\mathbf{r}_t$ be the pre-determined $\frac{M}{2}$-length DMRS sequences (as mentioned earlier in method-$1$) which  undergo $\frac{\pi}{2}$-phase rotation through diagonal matrix $P_1$ of sizes  $\frac{M}{2}\times\frac{M}{2}$.Where $P_1$ is a $\frac{M}{2}\times\frac{M}{2}$ diagonal matrix with diagonal entries given by $(e^{i(m \hspace{-5pt} \mod 2)\frac{\pi}{2}})$ this result in a $\frac{M}{2}$-length DMRS vector $\mathbf r_t^p= \mathbf P_1 \mathbf r_t$ with $\frac{\pi}{2}$-BPSK symbols. The spectrum shaping of the DMRS symbols is performed in time domain through a circular-convolution procedure with zero-padded $\mathbf{w}_t$ to produce a spectrum-shaped  DMRS sequences as,
  \begin{align}
 \mathbf{r}_t^s(n)&=\sum_{m=0}^{\frac{M}{2}-1}\mathbf r_t^p(n)\mathbf{w}_t\left((m+n)\hspace{-10pt}\mod\frac{M}{2}\right),\nonumber \\
 &\hspace{50pt} m, n \in \left[0,\frac{M}{2}-1\right]
 \end{align}
 
The spectrum-shaped  DMRS sequence is DFT precoded by means of  $\frac{M}{2}$-point DFT matrix as  $\mathbf{r}_f^s =\mathbf{D}_{\frac{M}{2}}\mathbf{r}_t^s$.  The DFT output of DMRS sequence generated above is mapped to \textit{port}-0 as}
\begin{equation*}
\begin{split}
     \mathbf {r}_f^{s_0}(k) &= \mathbf {r}_f^{s} \left(\frac{k}{2}\right) \hspace*{0.5 cm} k \in {0,2,4,....}\\
     &= 0 \hspace*{1.2 cm}\textnormal{otherwise}, 
\end{split}
\end{equation*}
and to \textit{port}-1 as 
\begin{equation*}
\begin{split}
\mathbf {r}_f^{s_1}(k)&= \mathbf {r}_f^{s} \left(\frac{k-1}{2}\right) \hspace*{0.5 cm}  k\in {1,3,5,....}\\
    &= 0 \hspace*{1.2 cm} \textnormal{otherwise}. 
 \end{split}
\end{equation*}
In the above equations, $\mathbf {r}_f^{s_0}$ and $\mathbf {r}_f^{s_1}$ indicate the frequency domain DMRS sequences on \textit{port}-0 and \textit{port}-1 respectively. It can be seen that with the proposed architecture the non-zero entries of DMRS sequence are exactly identical for both the ports i.e.,
 \begin{align}\label{eq:M2 Equivalence}
 \mathbf r_f^{s_0}(2k)&= \mathbf r_f^{s_1}(2k+1)\nonumber \\
                      &= \mathbf r_f^{p}(k)\mathbf w_f(k),
 \end{align}
 where $\mathbf r_f^{p}(k)$, $\mathbf w_f(k)$ are the $\frac{M}{2}$-DFT outputs of $\frac{\pi}{2}$-BPSK DMRS symbol $\mathbf{r}_t^p$,  filter $\mathbf{w}_t $ respectively. The DFT precoded spectrum-shaped data and DMRS vector of each port is mapped to a set of sub-carriers in frequency domain via a mapping matrix $M$ (described in Sec~\ref{sec:spectrum_shaping}). Finally, this mapped sequence is converted to time-domain via inverse-DFT operation as
\begin{align}
\mathbf{s}_t^0 &=\mathbf{D}_N^{\dagger}\mathbf{M}_f\mathbf{r}_f^{s_0} \nonumber \\
\mathbf{s}_t^1 &=\mathbf{D}_N^{\dagger}\mathbf{M}_f\mathbf{r}_f^{s_1}.
\end{align}
Using the above, the overall time-domain baseband signals for DMRS transmission i.e., $\mathbf{s}_t^0(t)$, $\mathbf{s}_t^1(t)$ with appropriate length cyclic prefix are generated as per equation (5.3.1) in 3GPP spec \cite{211}.
  		\begin{figure*}
	\centering
	\includegraphics[width=0.8\textwidth]{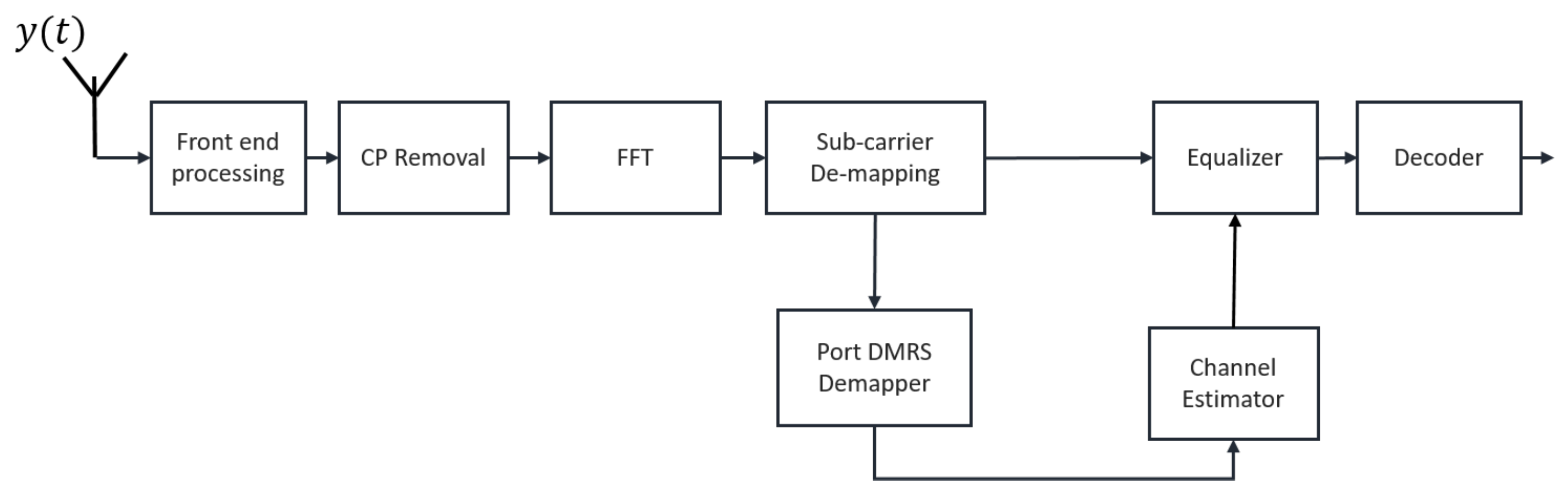}
	\caption{Base station receiver architecture for each receive antenna.}
	\label{fig:Rxer}
\end{figure*}
\subsection{\textbf{Summary of the transmission methods}}We presented two transmission methods for the data, DMRS waveform generation. Specifically, in method-$1$, the processing happens in frequency domain while in method-$2$ the processing happens in time domain via the circular-convolution operation. {Also, in method-$1$, a $M$-length DMRS sequence is spectrum shaped in frequency domain, whereas a length $\frac{M}{2}$ DMRS sequence is spectrum shaped in time domain. Irrespective of this difference, we show that both these methods are capable of estimating the channel perfectly}. Further using \eqref{eq:M1 Equivalence} and the DFT property that even indexed samples of $M$-point DFT of any arbitrary sequence will be identical to its $\frac{M}{2}$-point DFT output, it can be shown that
\begin{equation}\label{DFT Reln}
 \mathbf r_{f,M}^{p_0}(2k)=\mathbf r_{f,\frac{M}{2}}^{p_0}(k) \hspace{0.5cm} k=2l , \hspace{0.2cm} l=0,1,....\frac{M}{2}-1 
\end{equation}
where $\mathbf r_{f,M}^{p_0}$, $\mathbf r_{f,\frac{M}{2}}^{p_0}$ are $M$-point and $\frac{M}{2}$-point DFT outputs of $\mathbf{r}_t^p$ respectively.

Using \eqref{DFT Reln}, we can rewrite \eqref{eq:M1 Equivalence} as 
\begin{align*}
\mathbf{r}_f^{s_0}(2k)&= \mathbf{r}_f^{s_1}(2k+1) \nonumber \\
&= \mathbf {r}_{f,\frac{M}{2}}^{p_0}(k)\mathbf{w}_{f,\frac{M}{2}}(k)
\end{align*}
which is exactly identical to \eqref{eq:M2 Equivalence}.
Since input to IDFT is identical for both the methods, we conclude that the inverse DFT outputs of \textit{port}-0, \textit{port}-1 i.e., $\mathbf {r}_f^{s_0}$, $\mathbf {r}_f^{s_1}$ and the subsequent baseband signals generated through method-$1$ will be identical to that of generated using method-$2$. Using an example, we show in the Appendix that the channel estimation performance when these different transmitter methods are used will remain the same.

{\textbf{Remark:} \textit{The current 3GPP specifications for $5$G NR do not mention how the spectrum shaping and transmission for data and DMRS must be done for single as well as multiple antenna ports as it is left as an implementation choice. However, as we have shown extensively, this causes ambiguity at both the transmitter and receiver if not done in the right manner. Hence to avoid this ambiguity and also to avoid data loss on any antenna port, the designs mentioned above must be used.}}


     \section{Receiver Design}\label{sec:ReceiverDesign}
The receiver procedure explained next is common for both the transmission methods explained in previous sections. Hence, we do not distinguish between the transmission method-$1$ and method-$2$ in this section.

The receiver front end operations such as sampling, synchronization, CP removal and FFT are similar to a conventional DFT-s-OFDM-based system as shown in \ref{fig:Rxer}. Further, the ISI introduced by the propagation channel is assumed to be less than that of the CP length. Therefore, after CP removal and DFT, the data and DMRS signals on $k$th sub-carrier can be represented as (without loss of generality we consider only the initial M subcarriers of the DFT output, i.e., $k\in [0,M-1]$)
    \begin{align}
\mathbf y_d(k)&=\mathbf x_f^{s_0}(k)\mathbf h_{f,\mathtt{data}}^{0}(k)+\mathbf x_f^{s_1}(k)\mathbf h_{f,\mathtt{data}}^{1}(k)+\mathbf v(k)\nonumber \\
 \mathbf y_{\mathtt{DMRS}}^0(k)&=\mathbf r_f^{s_0}(k)\mathbf h_{f,\mathtt{DMRS}}^{0}(k)+\mathbf v_0(k) \label{eq:Port0 Rx} \\
\mathbf y_{\mathtt{DMRS}}^1(k)&=\mathbf r_f^{s_1}(k)\mathbf h_{f,\mathtt{DMRS}}^{1}(k)+\mathbf v_1(k) \label{eq:Port1 Rx}.
  \end{align}
In the above, $\mathbf y_d$ correspond to the received data vector with data symbols from both the ports (recall that $2$-antenna ports can support $2$-MIMO stream transmissions). $\mathbf y_{\mathtt{DMRS}}^0$, $\mathbf y_{\mathtt{DMRS}}^1$ correspond to the received DMRS vectors on \textit{port}-0 and \textit{port}-1 respectively. $\mathbf h_{f,\mathtt{DMRS}}^{0}=\mathbf{D}_M\mathbf{h}_t^0$, $\mathbf h_{f,\mathtt{DMRS}}^{1}=\mathbf{D}_M\mathbf{h}_t^1$ correspond to frequency response of the time-domain wireless channel impulse response $\mathbf{h}_t^0$ on \textit{port}-0 and $\mathbf{h}_t^1$ on \textit{port}-1 respectively and $\mathbf x_f^{s_0}$, $\mathbf x_f^{s_1}$, $\mathbf r_f^{s_0}$, $\mathbf r_f^{s_1}$ are the transmitted data and DMRS sequences defined in Section~\ref{sec:SignalModel}. The noise vectors $\mathbf{v}, \mathbf{v}_0$ and $\mathbf{v}_1$ are \textit{i.i.d.} complex Gaussian random variables with zero-mean and co-variance $\sigma^2\mathbf{I}$ where $\mathbf{I}$ is an identity matrix and $\sigma^2$ is a constant indicating the variance of each noise sample.

\begin{figure}
	
	\centering
	\includegraphics[width=0.5\columnwidth,height=5cm]{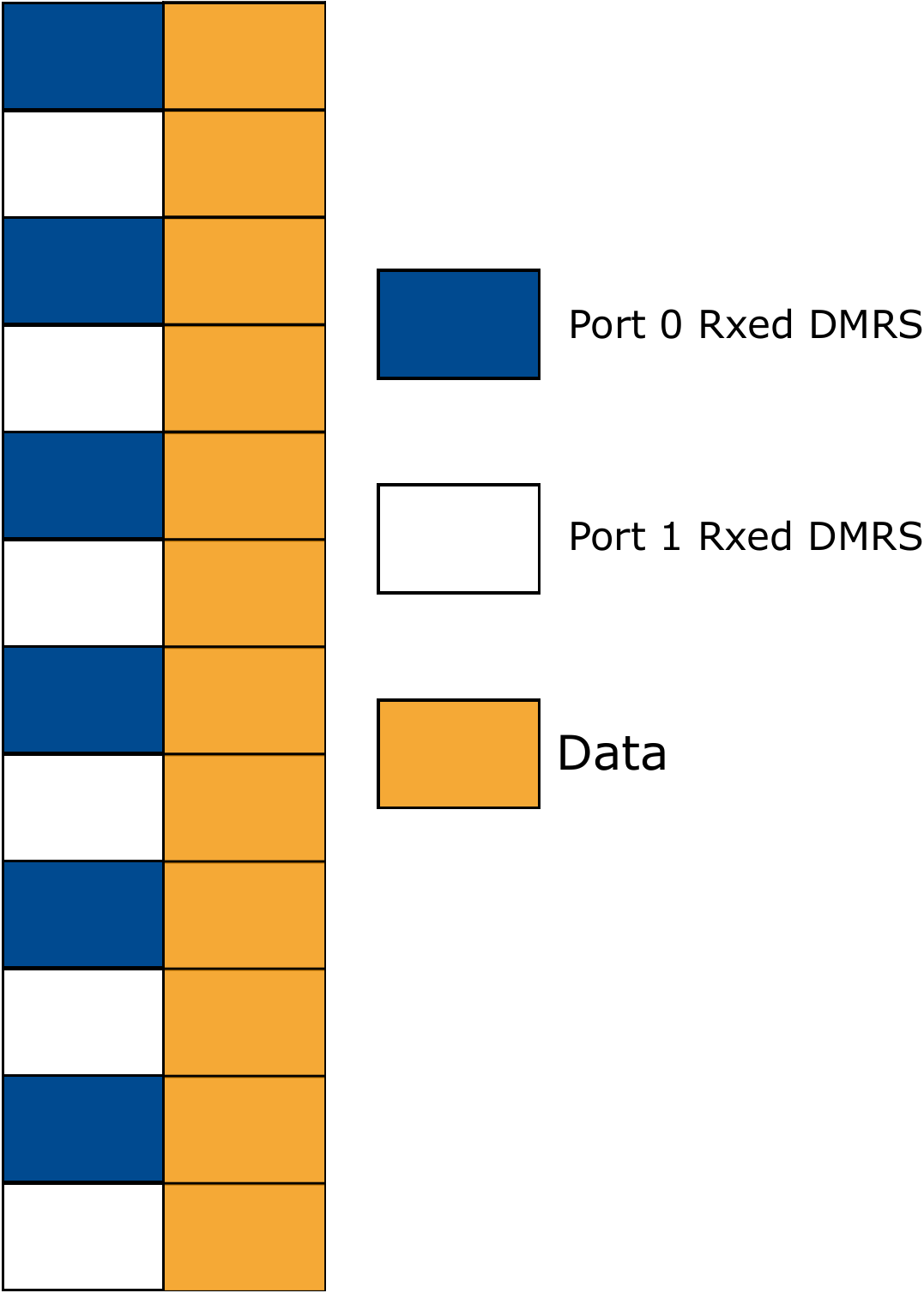}
	\caption{DMRS and data symbols in a OFDM resource grid.}
	\label{fig:Data_symb}
	\vspace{-12pt}
\end{figure}

In practice, for low to medium user speeds the time variations of the multipath channel across consecutive OFDM symbols as shown in Fig.~\ref{fig:Data_symb} will be minimal and hence without loss of generality we consider that
\begin{equation*}
\mathbf h_{f,\mathtt{DMRS}}^0(k)=\mathbf h_{f,\mathtt{data}}^0(k) .
\end{equation*} 
\begin{equation*}
\mathbf h_{f,\mathtt{DMRS}}^1(k)=\mathbf h_{f,\mathtt{data}}^1(k) .
\end{equation*}

This is a common assumption made in the design of $4$G and $5$G cellular systems. 

    \subsubsection{\textbf{Channel estimation}}
As per $3$GPP specifications, the spectrum shaping filter $\mathbf{w}_t$ is implementation-specific i.e., different UEs can use different filters based on their hardware implementation and hence the exact filter being used is unknown at the base station receiver \cite{101}, \cite{213}. Hence, the channel estimation module at the receiver should now estimate the impulse response of filter and wireless channel jointly. In our work, we use a DFT-based channel estimation technique to estimate the joint channel impulse response for the $M$ allocated sub-carriers. A simple least-squares based technique with tone averaging or linear interpolation  based on assumption that the channel is constant across consecutive sub-carriers does not work well in this case due to the presence of the spectrum shaping filter, because spectrum shaping considerably changes channel across consecutive sub-carriers based on the shape of the filter shown in Fig.~\ref{fig:spectrum_shaping}.
    
As already mentioned in Section~\ref{sec:SignalModel}, a $M$-length data vector will be associated with  $\frac{M}{2}$-length DMRS vector. Firstly, we show that  a $M$-length frequency domain channel vector (as the data allocation is $M$, the channel on all of these $M$ tones must be estimated for demodulation) corresponding to $M$-length data symbol can be perfectly constructed from  $\frac{M}{2}$-length DMRS sequence for both ports.
\subsubsection{\textbf{Channel estimation on \textit{port}-0}}\label{sec:chan_est_length_port}
As mentioned earlier, \textit{port}-0 carries DMRS only on even numbered sub-carriers which are extracted and expressed in terms of  $\frac{\pi}{2}$-BPSK DMRS as follows 
\begin{align}
\tilde{ \mathbf y}_{\mathtt{DMRS}}(k) & = \mathbf y^0_{\mathtt{DMRS}}(2k),    k=0,1,..\frac{M}{2}-1 \nonumber \\
 &=\mathbf r_f^{s_0}(2k)\mathbf h_{f,\mathtt{DMRS}}^0(2k)+\mathbf v_0(2k), \label{eq:tempe1}\\  
  &=\mathbf {r}_f(2k)\mathbf {w}_f(2k)\mathbf h_{f,\mathtt{DMRS}}^0(2k)+\mathbf{v}_0(2k), \label{eq:tempe2}
 \end{align}
where \eqref{eq:tempe1} results from \eqref{eq:Port0 Rx}, and \eqref{eq:tempe2} results from \eqref{eq:M1 Equivalence}. Invoking the equivalence between $M$-point DFT and $\frac{M}{2}$-point DFTs \eqref{DFT Reln}, the above equation can be represented as
\begin{equation}
	\tilde{ \mathbf y}_{\mathtt{DMRS}}^0(k)=\left[\mathbf{D}_{\frac{M}{2}} \mathbf{\tilde{r}}_t^p\right] \left[\mathbf{D}_{\frac{M}{2}}\left(\mathbf w_t \odot \mathbf  h_{t,\mathtt{DMRS}}^0\right)\right]+\mathbf{v}_0
\end{equation}
 where $\odot$ indicates the circular-convolution operation, $\mathbf {h}_{t,\mathtt{DMRS}}^0$ is the impulse response of the wireless channel on \textit{port}-0, $\mathbf{\tilde{r}}_t^p(n)$ is defined in section \ref{sec:SignalModel} 

We perform channel estimation on ${\tilde{\mathbf {y}}_{\mathtt{DMRS}}}$ as follows - we first perform a least squares based channel estimation and then on the resulting output we take an $\frac{M}{2}$-point IDFT. This gives the joint impulse response of filter and  the wireless channel as -
\begin{align}
D_{\frac{M}{2}}^\dagger \left(\frac{\mathbf {\tilde{y}_{\mathtt{DMRS}}}}{\mathbf{D}_{\frac{M}{2}} \mathbf{\tilde{r}}_t^p }\right)&=\underbrace{\mathbf w_t(n)\odot \mathbf  h_{\mathtt{f,DMRS}}^0(n)}_{\mathbf h_{\mathtt{eff}}}. \label{step1:channelest}
\end{align}
The length of $\mathbf h_{\mathtt{eff}}$ will be $\max\left(\mathtt{length}(\mathbf{w}_t),  \mathtt{length}(\mathbf  h_{\mathtt{t,DMRS}}^0)\right)$. Irrespective of the pulse shaping filter, the reference signal design should ensure that the DMRS sequence length will be at-least twice that of the impulse response of the wireless channel i.e., the length of $\mathbf  h_{\mathtt{t,DMRS}}^0)$ is assumed to be less than $\frac{M}{2}$ \cite{CIR1} which is typically the case for practical wireless channel models considered by $3$GPP \cite{901}. From the above we conclude that $\mathbf h_{\mathtt{eff}}$ completely captures the joint impulse response of the spectrum shaping filter and also the wireless channel. 
   
A de-noising time domain filter \cite{Denoise} is then applied to reduce noise in ~\eqref{step1:channelest}. This filter $\mathbf {f}(n)$ is defined as 
 \begin{equation*}
 \begin{split}
 \mathbf {f}(n)&=1, \hspace{0.6cm} 0\leq n \leq f_c-1 ,M-f_c\leq n\leq 1 \\ 
 	&=0, \hspace*{0.6cm} \textnormal{otherwise}\\
 	\end{split}
 \end{equation*}
 where $f_c$ is the ``cut-off'' point of the time domain filter which is commonly chosen as the length of the wireless channel length $\mathtt{length}(\mathbf  h_{\mathtt{t,DMRS}}^0$ if it is known \emph{apriori} or it is set to the cyclic prefix length in case no knowledge about the wireless channel is available. The rest of the samples are set to $0$. This filter helps to extract only the useful samples of the CIR while reducing the noise in the rest of the samples. For more details, please see \cite{Denoise}. The effective impulse response after de-noising is given as 
  \begin{equation*}
 \hat{\mathbf{h}}_{\mathtt{eff}}(n)=\mathbf {h}_{\mathtt{eff}}(n)\mathbf {f}(n), \hspace*{0.5cm} 0\leq n\leq M-1
 \end{equation*}
 Lastly, the time domain filtered samples are transformed via a $M$-point DFT to recover the frequency-domain channel estimates on each sub-carrier $k\in[0,M-1]$  as $\hat{\mathbf{h}^f}_{\mathtt{eff}}=\mathbf{D}_M\hat{\mathbf h}_{\mathtt{eff}}(n)$. This can be further used for \textit{port}-0 data demodulation using well-known techniques.

  \subsubsection{\textbf{Channel estimation on \textit{port}-1}} 
\begin{figure}	
	\centering
	\includegraphics[width=0.8\columnwidth]{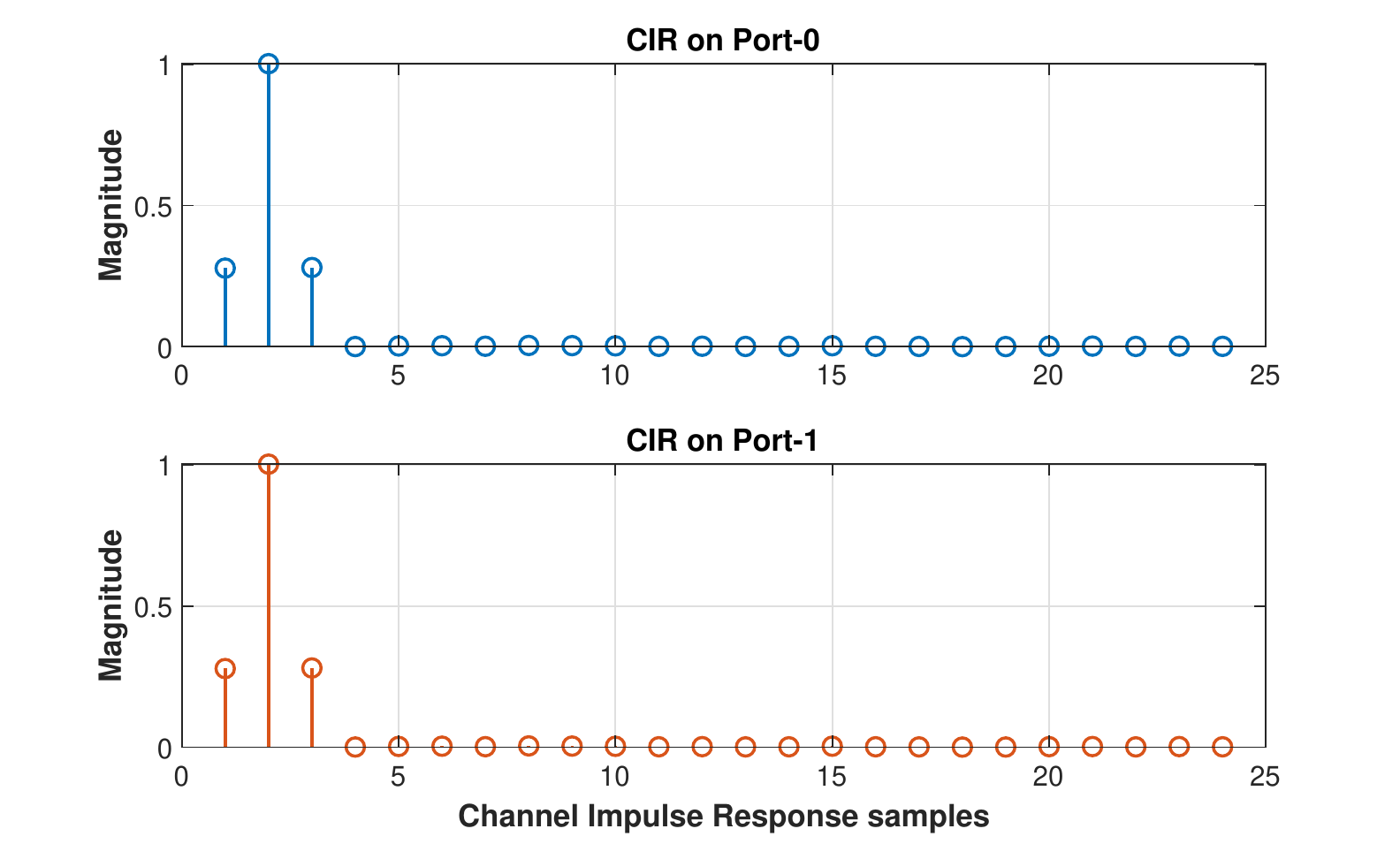}
	\caption{Magnitude of the estimated channel impulse response on \textit{port}-0 and \textit{port}-1.}
	\label{fig:CIR}
	\vspace{-5pt}
\end{figure}

As mentioned earlier, \textit{port}-1 carries DMRS only on odd sub-carriers which are extracted and expressed as follows 
    \begin{equation*}
    \tilde{ \mathbf y}_{\mathtt{DMRS}}^1(k) = \mathbf y^1_{\mathtt{DMRS}}(2k+1),    k=0,1,\ldots, \frac{M}{2}-1
      \end{equation*}
Using \eqref{eq:Port1 Rx}, the above equation can be written as 
    \begin{equation}
  \tilde{ \mathbf y}_{\mathtt{DMRS}}^1(k)  =\mathbf r_f^{s_1}(2k+1)\mathbf h_{f,\mathtt{DMRS}}^1(2k+1)+\mathbf v_1(2k+1) \label{eq:y1DMRS}
    \end{equation}
Assuming that the wireless channel remains constant across consecutive sub-carriers (again a common assumption in $3$GPP designs), we have
 \begin{equation}\label{Approx}
 	\mathbf {h}_{f,\mathtt{DMRS}}^1(2k+1)\approx \mathbf {h}_{f,\mathtt{DMRS}}^1(2k).
 \end{equation}
Using \eqref{eq:M1 Equivalence} and \eqref{Approx}, \eqref{eq:y1DMRS} can be expressed as
  \begin{align}
 \tilde{ \mathbf y}_{\mathtt{DMRS}}^1(k)&=\mathbf {r}_f(2k)\mathbf {w}_f(2k)\mathbf h_{f,\mathtt{DMRS}}^1(2k)+\mathbf v_1(2k+1)\nonumber \\
&=\left[\mathbf{D}_{\frac{M}{2}} \mathbf{\tilde{r}}_t^p\right] \left[\mathbf{D}_{\frac{M}{2}}\left(\mathbf w_t \odot \mathbf  h_{t,\mathtt{DMRS}}^1\right)\right]+\mathbf{v}_1
  \end{align}
Further processing steps such as the least-squares based channel estimation, de-noising and transforming the effective impulse response to frequency domain are identical to the procedure followed for channel estimation on \textit{port}-0.
For the case of AWGN channel i.e., 
	\begin{equation*}
	\mathbf {h}_{f,\mathtt{DMRS}}^0(k)=\mathbf {h}_{f,\mathtt{DMRS}}^1(k)=1\hspace{0.2cm} \forall k,
	\end{equation*}
the estimated joint impulse response ${\mathbf h_{\mathtt{eff}}}$ on \textit{port}-0 and \textit{port}-1 is shown in Fig.~\ref{fig:CIR}. It can be noticed that the estimated impulse response is identical for both the ports.

 \subsubsection{\textbf{Equalization and data demodulation}:} The estimated channel on \textit{port}-0 and \textit{port}-1 will be employed for channel equalization of data streams. Specifically, we construct an MMSE-equalization filter using the channel estimates obtained previously and then equalize the received signal samples on all the receive antennas of the base station. 
The equalized data streams are demodulated to generate soft log-likelihood ratio values which are given as input to the channel decoder module for further bit-level processing.

\section{Numerical Results}\label{sec:Numerical Results}
In this section, we present various numerical results that show
\begin{itemize}
	\item The PAPR comparison between the $\frac{\pi}{2}$-BPSK based  DMRS sequences and the existing $3$GPP ZC-based DMRS sequences.
	\item Link level block error rate (BLER) comparison for the data transmissions employing $\frac{\pi}{2}$-BPSK based DMRS sequences and existing $3$GPP ZC-based DMRS sequences for various sequence lengths and various bandwidth allocations.
	\item BLER performance for the data transmissions on \textit{port}-0 and \textit{port}-1 in the case of MIMO two stream transmissions.	
\end{itemize}
Unless otherwise mentioned, the simulation assumptions shown in Table~\ref{tab:Sim Params} are used throughout this paper.

 \begin{table}[t]
	\centering
	{
		\caption{Simulation Assumptions for BLER comparisons}
		\label{tab:Sim Params}
		\renewcommand{\arraystretch}{1.5}
		{\fontsize{7}{7}\selectfont
			\begin{tabular}{|c|c|} \hline 
				\textbf{Parameter} & \textbf{Value} \\ 
				
				\hline \hline
							Channel Type & PUSCH\\
				\hline
				System Bandwidth & $20$ MHz\\
				\hline
				Sub-carrier spacing & $15$ KHz \\
				\hline
			   Allocated PRBs & $1$-$16$ PRBs\\
				\hline
				Channel Model & TDL-C $300$ns \\
				\hline
			
				Number of UE transmitter antennas & $1$\\
				\hline
				Number of UEs  & $1$, $2$\\
\hline				

				Number of BS receiver antennas & $2$, $4$\\
\hline
				
				Number of MIMO streams & $1$, $2$ \\
				\hline
				
				Channel Coding & $3$GPP NR LDPC\\
				\hline
			
				Equalizer & MMSE\\
				\hline

			\end{tabular}
		}
	}
\end{table}

\begin{figure}	
	\centering
	\includegraphics[width=0.8\columnwidth]{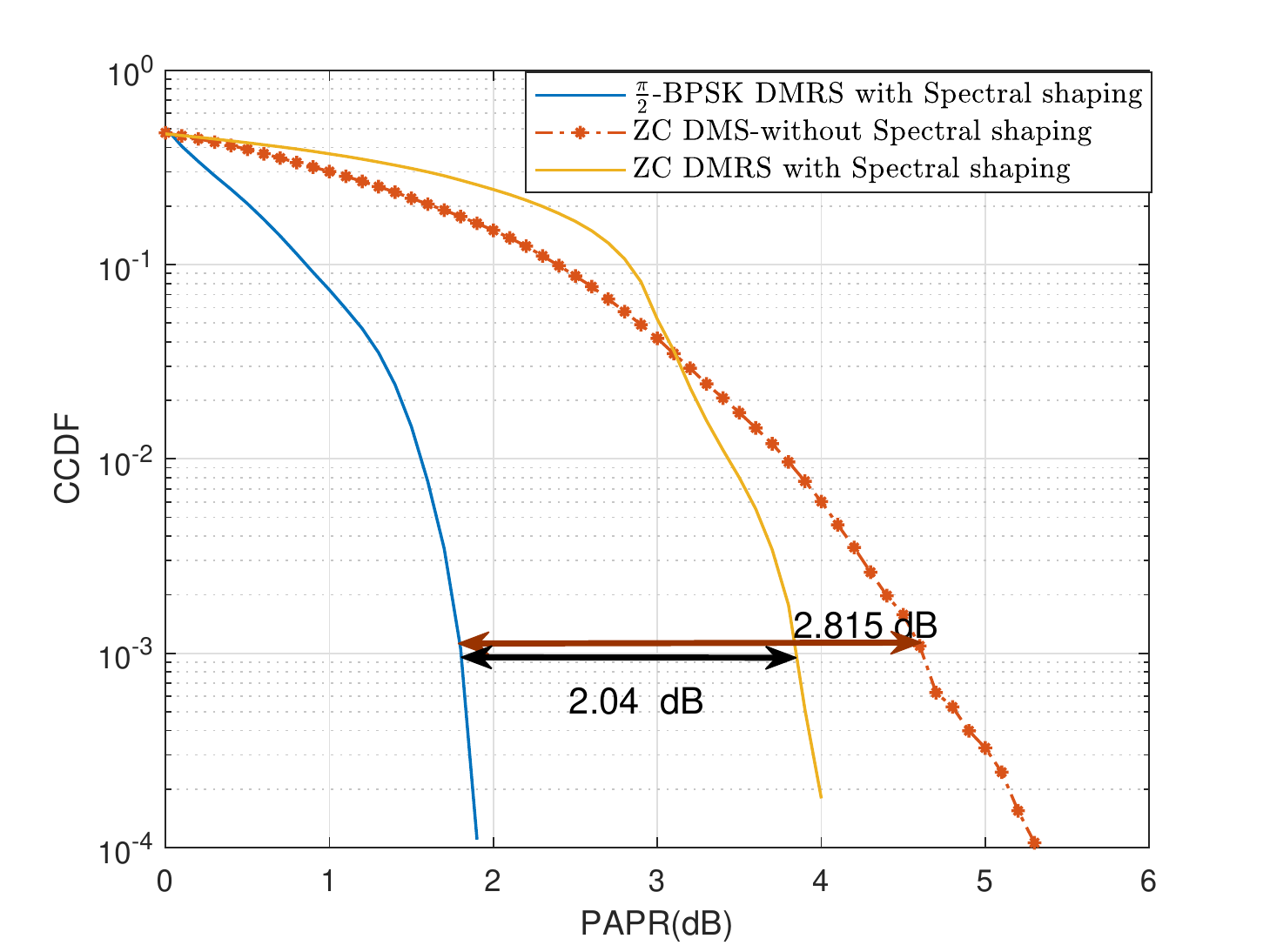}
	\caption{PAPR of length $96$ ZC and $\frac{\pi}{2}$-BPSK DMRS sequences.}
	\label{fig:PAPR22}
	\vspace{-10pt}
\end{figure}

The CCDF of PAPR for ZC and $\frac{\pi}{2}$- BPSK sequences is shown in Fig.~\ref{fig:PAPR22}. The ZC sequences considered in this case are as defined in \cite[section 5.2.2]{211} with length $96$. The PAPR of both with and without spectrum shaping of ZC sequences is shown in the figure. As can be seen from the figure, the  $3$GPP ZC sequences without spectrum shaping have a PAPR (at the $10^{-3}$ CDF point) that is $2.8$dB more than the $\frac{\pi}{2}$-BPSK sequences. When spectrum shaping is applied to the ZC-DMRS, the PAPR is slightly reduced from that of un-filtered ZC sequences. However, the PAPR of the filtered ZC sequence is still $2.0$ dB larger than the PAPR of the $\frac{\pi}{2}$ BPSK sequences with the same spectrum shaping. Moreover, as we increase the number of allocated sub-carriers for data transmission, the PAPR gap between $3$GPP ZC sequence and  $\frac{\pi}{2}$ BPSK increases even further.

\begin{figure}
	
	\centering
	\includegraphics[width=0.8\columnwidth]{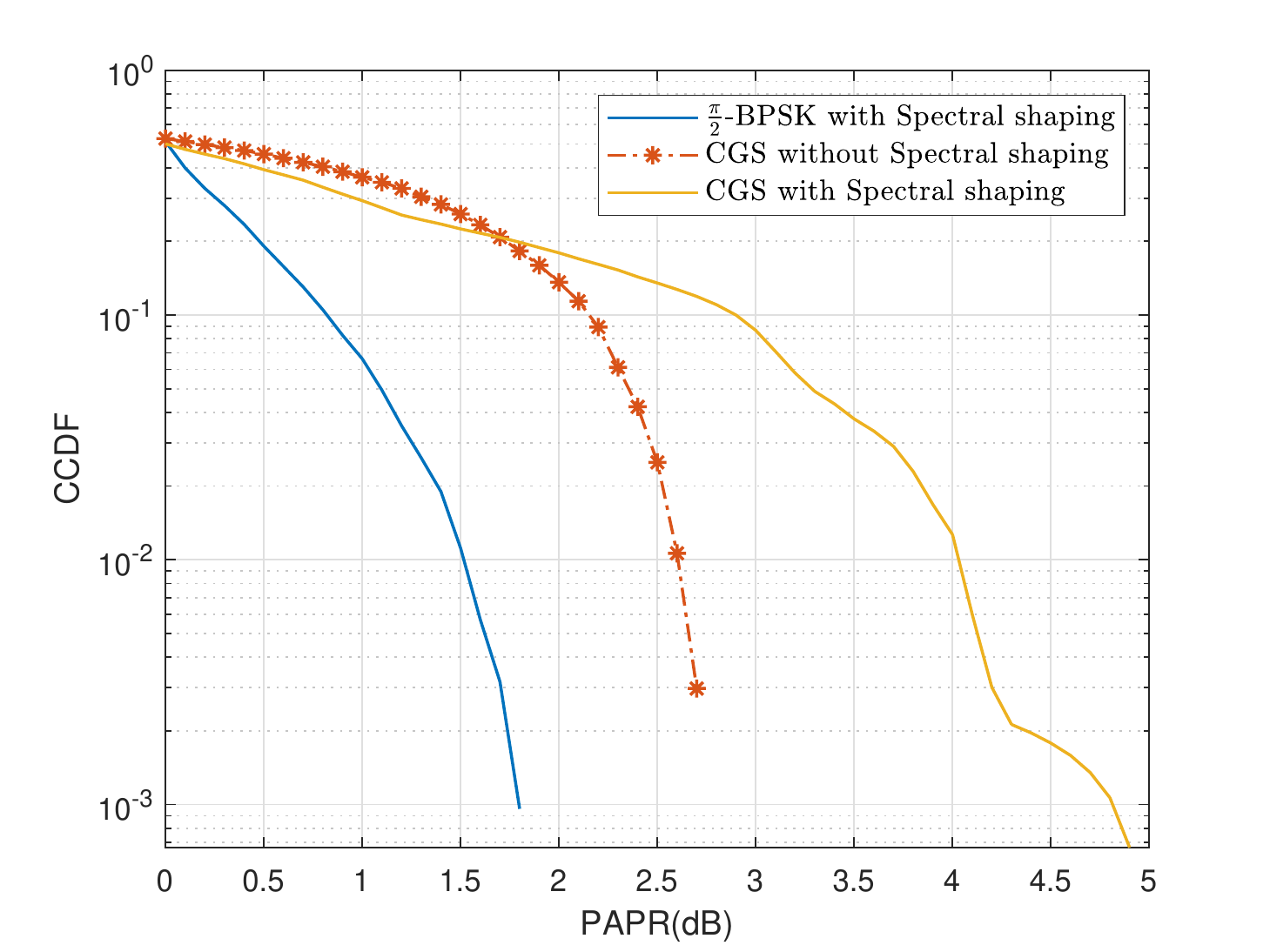}
	\caption{PAPR of  length-12 $3$GPP CGS and $\frac{\pi}{2}$-BPSK DMRS sequences  }
	\label{fig:PAPR_CGS}
	\vspace{-12pt}
\end{figure}

The CCDF of PAPR for ZC and $\frac{\pi}{2}$-BPSK for smaller lengths $(N=12)$ is shown in Fig.~\ref{fig:PAPR_CGS}. As discussed in section \ref{sec:SignalModel}, for smaller lengths $(N<30)$, $3$GPP employs computer generated sequences (CGS)as DMRS. It can be seen from the figure that the PAPR of the spectrum shaped CGS sequences is almost $1.2$ dB larger than the PAPR of the $\frac{\pi}{2}$ BPSK sequences. Moreover, it can also be noticed that for CGS the PAPR is further increased with  filtering. Hence, these results conclude that the $\frac{\pi}{2}$ BPSK sequences designed in \cite{Qual} are far superior compared to the existing sequences in improving the cell-coverage.

\begin{figure}
	\centering
	\includegraphics[width=0.8\columnwidth,height=5.5cm]{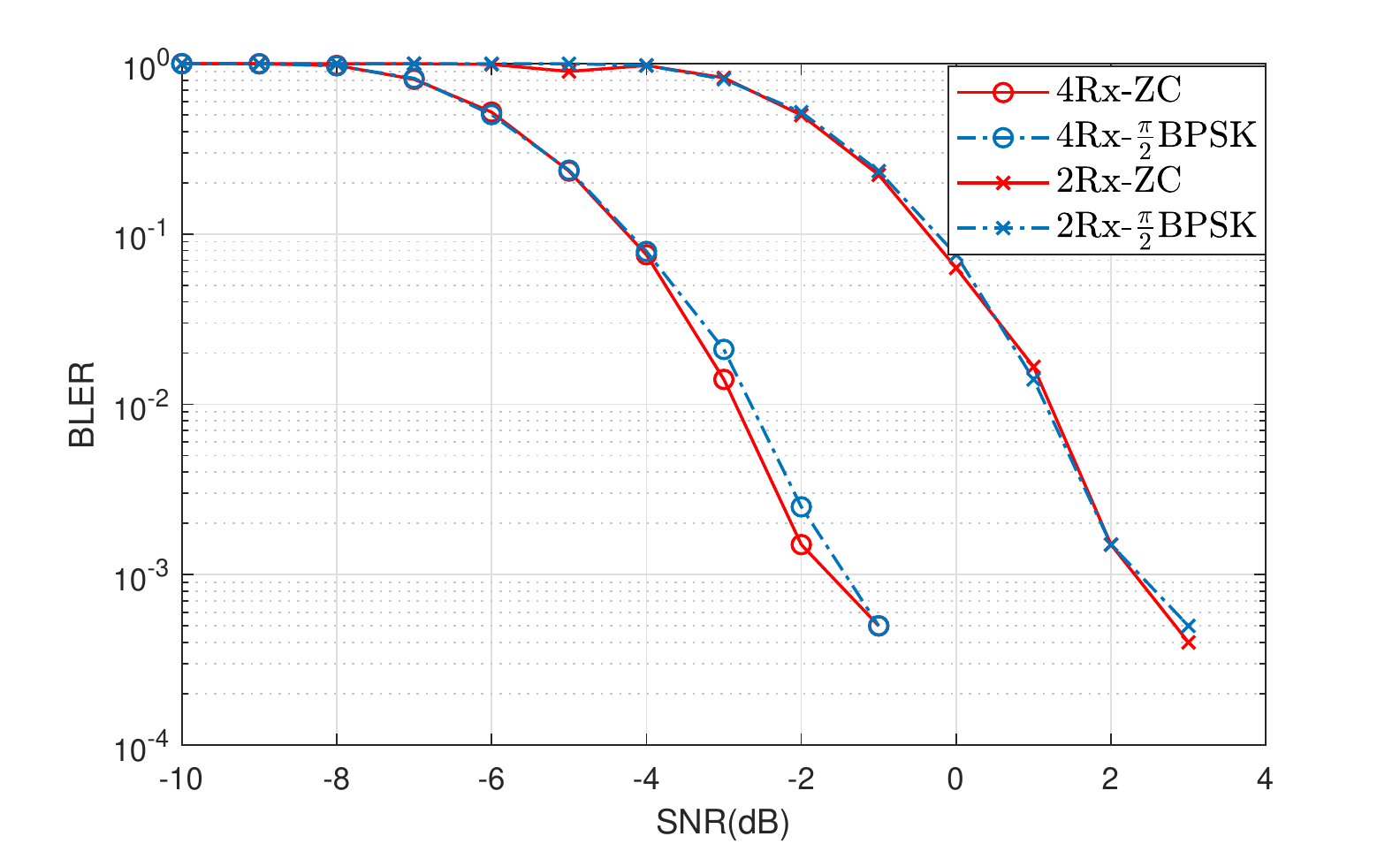}
	\caption{BLER comparison of length-$24$ ZC and $\frac{\pi}{2}$-BPSK DMRS sequences.}
	\label{fig:BLER_24}
	\vspace{-12pt}
\end{figure}

The block error rate performance for a single stream PUSCH transmission in shown in Fig.~\ref{fig:BLER_24}. Here, DMRS is transmitted on \textit{port}-0. Note that ZC sequences are used for comparing the BLER performance because these sequences have a power  density which is frequency-flat and hence treats every sub-carrier equally and can estimate the channel equally well across the entire bandwidth. Hence, the goal for the newly designed sequences is to ensure that they match the performance of these ZC sequences. In this figure, the results are shown for the cases when the base station receiver employs $2$ and $4$ receive antennas. {From Fig.~\ref{fig:BLER_24}, it can be clearly seen that irrespective of number of receive antennas, the link level performance of $\frac{\pi}{2}$-BPSK DMRS is equivalent to that of $3$GPP ZC-sequences although the newly designed sequences are not frequency-flat like ZC}. 

\begin{figure}
	\centering
	\includegraphics[width=0.8\columnwidth]{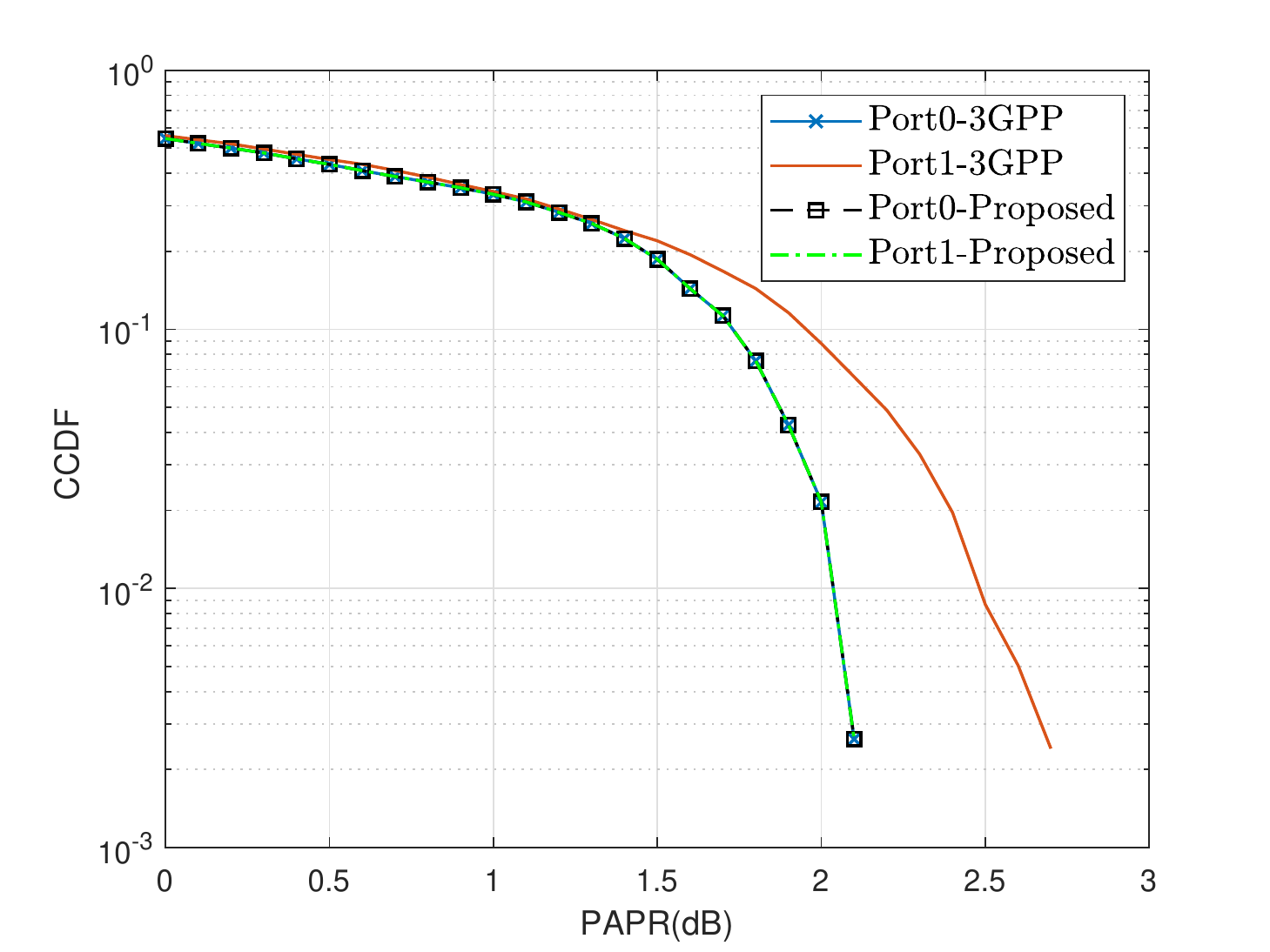}
	\caption{PAPR comparison of length-$6$ DMRS sequences on \textit{port}-0 and \textit{port}-1 with proposed and $3$GPP transmitter designs.}
		\vspace{-10pt}
	\label{fig:PAPR_our_3gpp}
\end{figure}

\begin{figure}
	\centering
	\includegraphics[width=0.8\columnwidth]{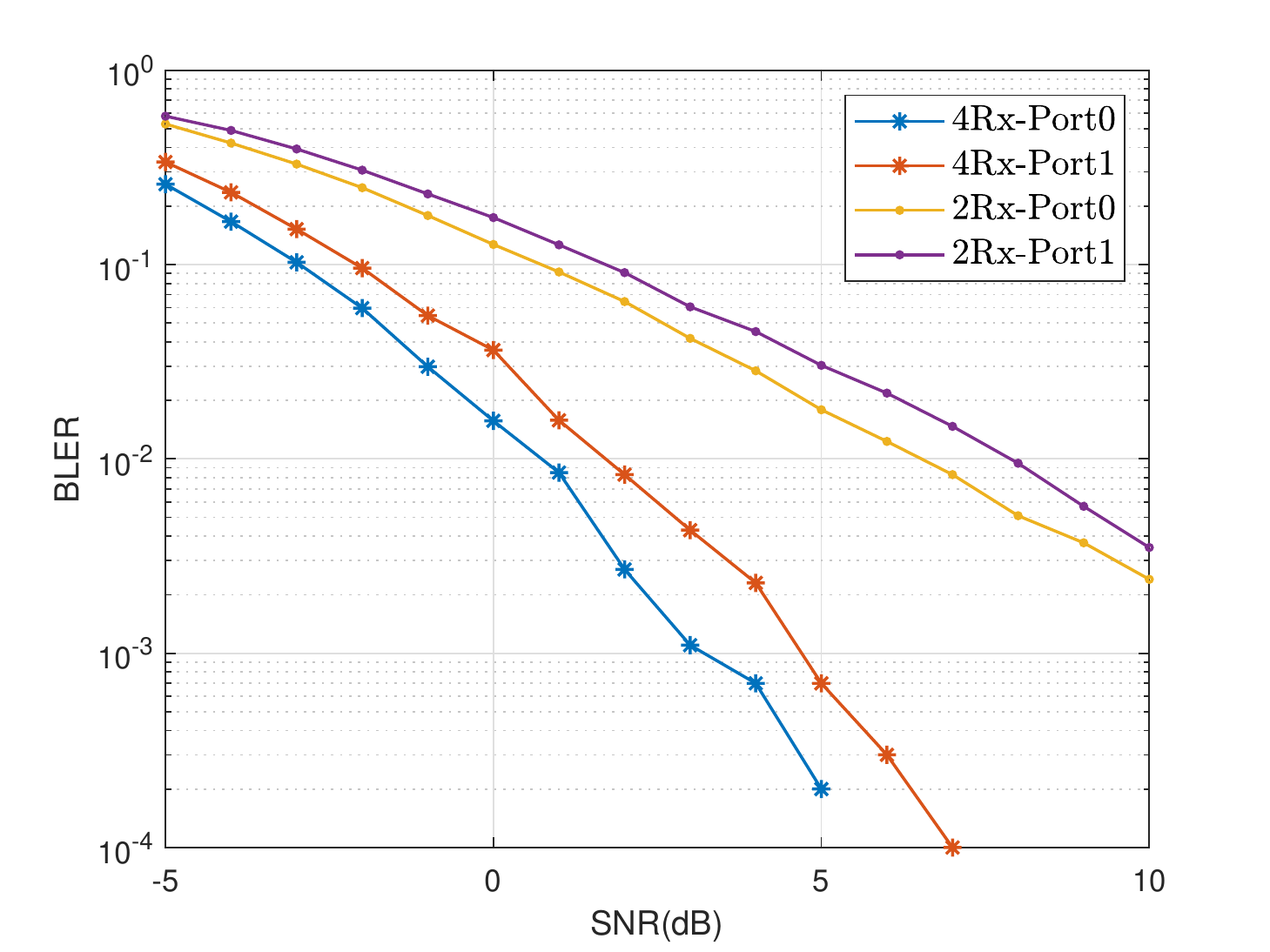}
	\caption{BLER comparison of length-$6$ DMRS sequences on \textit{port}-0 and \textit{port}-1 with   $3$GPP transmitter design.}
		\vspace{-10pt}
	\label{fig:TDL_2_4_3gpp}
\end{figure}

We next consider the performance of the proposed transmitter designs for the case of two MIMO streams transmission setting where DMRS is transmitted on both \textit{port}-0 and \textit{port}-1. Firstly, we show the drawbacks of the existing design in $3$GPP in Figs.~\ref{fig:PAPR_our_3gpp} and \ref{fig:TDL_2_4_3gpp}. It is seen that when the $3$GPP transceiver is used, there is a clear difference in the performance both in terms of PAPR and BLER across \textit{port}-0 and \textit{port}-1. This is highly undesirable as the data on two different ports will behave differently and practically \textit{port}-1 is useless. This problem is addressed using our proposed transceiver design as claimed earlier. We next show that it is indeed the case.

\begin{figure}
		\centering
	\includegraphics[width=0.8\columnwidth]{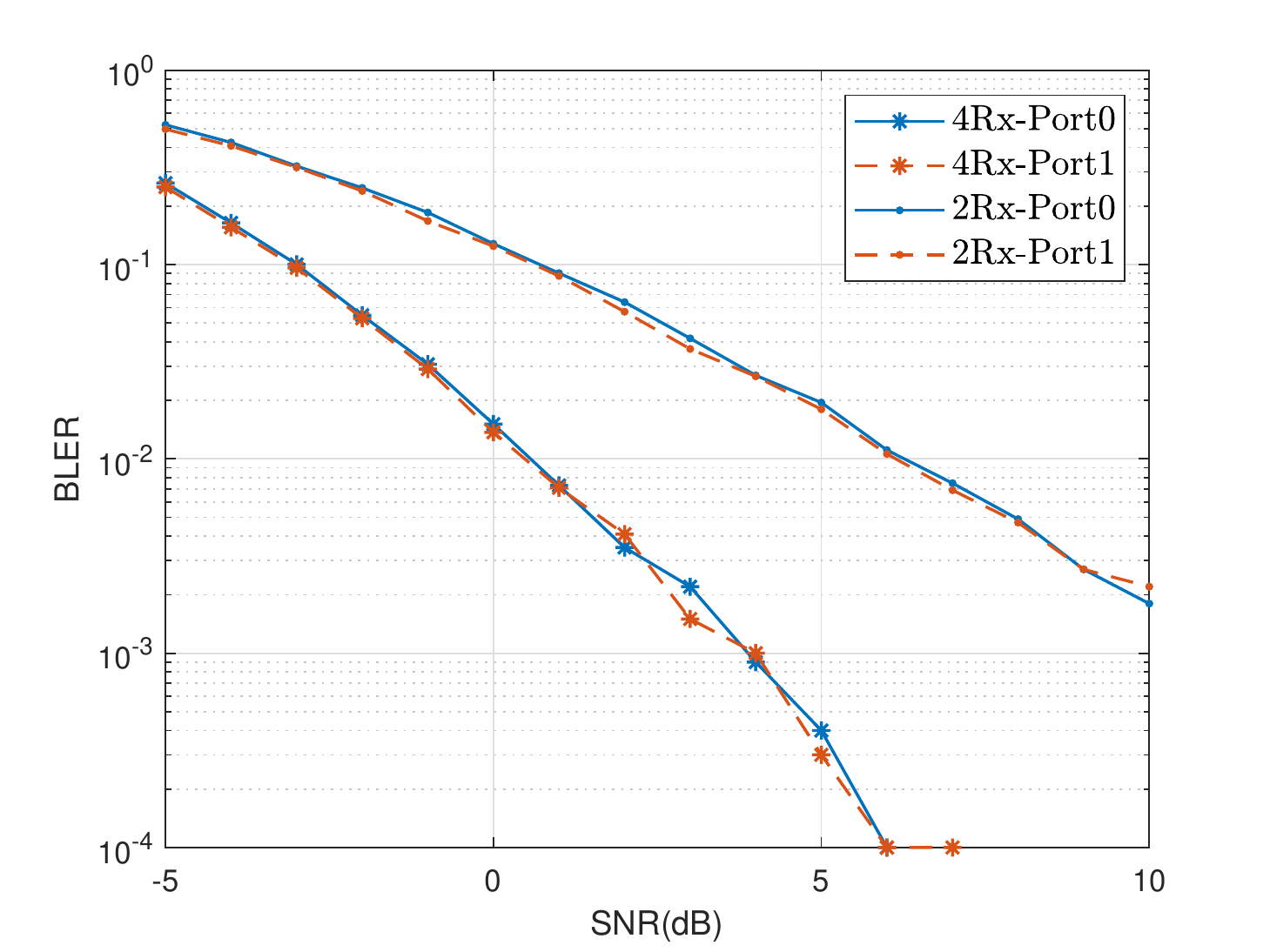}
	\caption{BLER comparison of length-$6$ DMRS sequences on \textit{port}-0 and \textit{port}-1 with proposed transmitter design.}
		\vspace{-10pt}
	\label{fig:TDL_2_4_our}
\end{figure}

In Figs.~\ref{fig:PAPR_our_3gpp} and \ref{fig:TDL_2_4_our}, we show the PAPR and BLER performance for the two MIMO streams transmission setting where DMRS is transmitted on both \textit{port}-0 and \textit{port}-1 using our proposed method-$1$ transceiver design. It can be seen that both the PAPR as well as the BLER is identical for both the ports confirming that the proposed transmitter design produces identical DMRS sequences on both the ports.

\begin{figure}
	\centering
	\includegraphics[width=0.8\columnwidth]{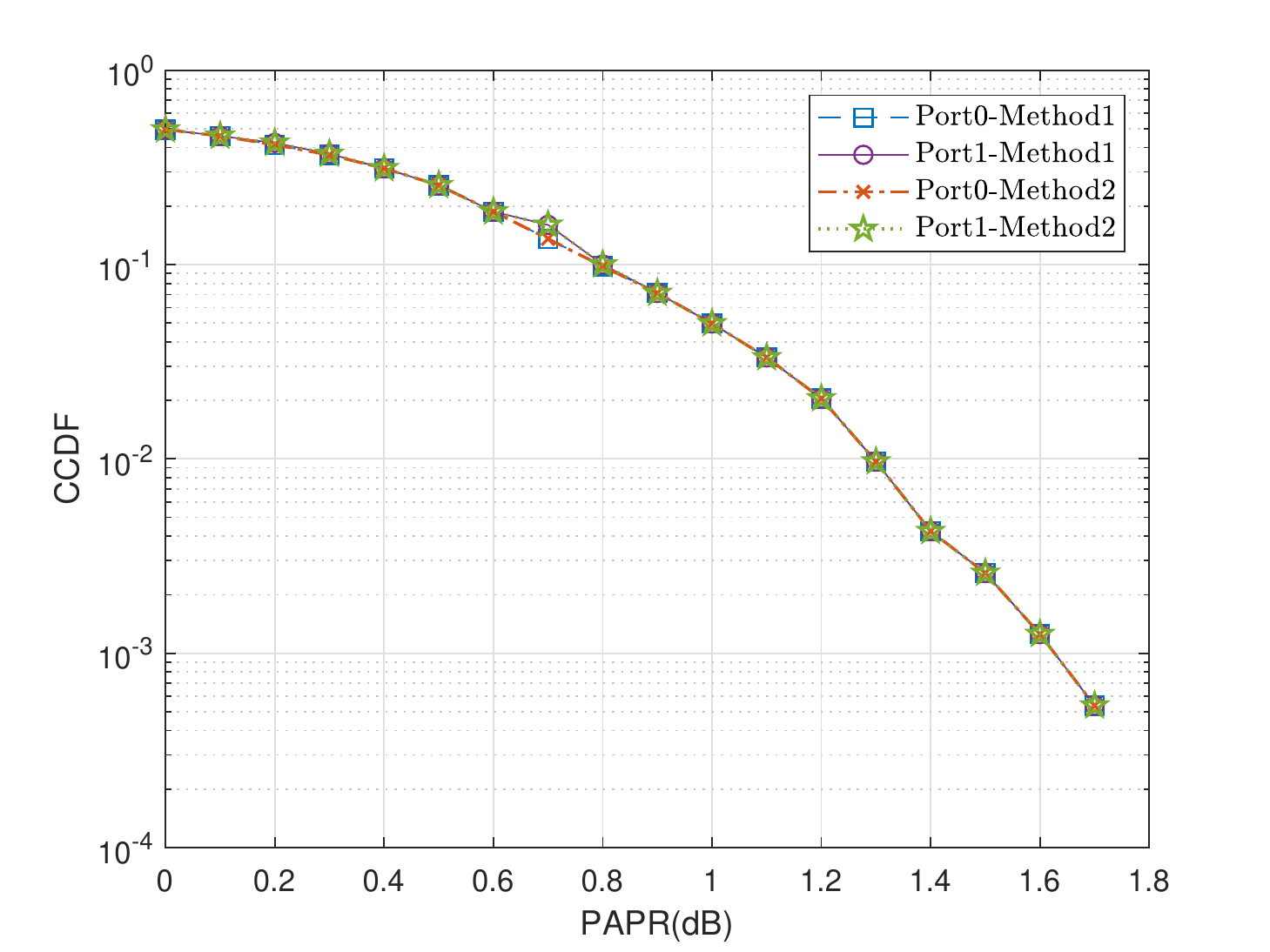}
	\caption{ PAPR comparison of length-$12$ DMRS sequences on \textit{port}-0 and \textit{port}-1 with method-$1$ and method-$2$ transmitter designs respectively.}
	\label{fig:M1-M2}
\end{figure}

In Fig.~\ref{fig:M1-M2}, PAPR of the DMRS sequences on \textit{port}-0 and \textit{port}-1 generated by method-$1$ and method-$2$ is shown. It can be seen that PAPR is exactly same for both \textit{port}-0 and \textit{port}-1 in both the methods confirming that proposed transmitter designs are equivalent. The same is the case with BLER performance as well. Therefore, the proposed methods~$1$ and $2$ have shown to be equivalent both analytically and numerically.

\section{Conclusion}\label{sec:Conclusion}
In this paper we proposed a low PAPR reference signal transceiver design for $3$GPP $5$G NR $\frac{\pi}{2}$-BPSK based uplink transmissions. Using the proposed design, the PAPR of the reference signal is significantly minimized compared to the current design of Rel-15 5G NR systems. Such a design considerably helps to improve the coverage of the $5$G systems. Specifically, we have shown a frequency domain and a time domain transceiver design both of which are equivalent and result in same system performance in terms of PAPR and also BLER. We have shown how the proposed design can be extended to the case of a MIMO transmission without causing any discrepancy on different MIMO streams which is not the case for the current Rel-$15$ $3$GPP $5$G NR uplink design.

\section*{Appendix}\label{sec:Appendix}
In this section we present the spectrum shaped DFT outputs of \textit{port}-0 and \textit{port}-1 generated using the proposed transmitter design. In Table~\ref{tab:Method1_CIR} we present the effective channel impulse response (CIR) estimated from both the ports in a noise-less scenario. We consider a $3$-tap spectrum shaping filter with impulse response $w_t=[-0.28~1~-0.28]$. For convenience we consider a flat and identical wireless channel for both DMRS ports, i.e., $\mathbf h_{f,\mathtt{DMRS}}^0(k)=\mathbf h_{f,\mathtt{DMRS}}^1(k)=1~~\forall k\in \left[0,M-1\right]$. Let $r_t=[1 1 1 0 1 1]$ be a $6$-length low PAPR BPSK-based DMRS sequence corresponding to length-$12$ data allocation. This sequence is passed through the transmitter design as shown in Fig.~\ref{fig:DMRS_M1} and Fig.~\ref{fig:DMRS_M1_p1} or Fig.~\ref{fig:M2_DMRS} corresponding to the method-$1$ or method-$2$ transmitter designs. The resulting output is shown in Table~\ref{tab:Method1_output}. As proved earlier, it can be seen from this table that the non-zero entries of the DMRS sequences on both the antenna ports are the same. 
\begin{table}[H]
	\centering
	{
		\caption{ DMRS output on \textit{port}-0, \textit{port}-1}
		\label{tab:Method1_output}
		\renewcommand{\arraystretch}{1.5}
		{\fontsize{8}{8}\selectfont
			\begin{tabular}{|c|c|c|} \hline 
			 \textbf {Subcarrrier Index}	&\textbf {\textit{port}-0 } &\textbf {\textit{port}-1 } \\
				
				\hline \hline
				$0$ & $-0.6223-1.2445i$&$ 0.000+0.000i$   \\
				\hline
				$1$ &$ 0.000+0.000i$	&$-0.6223-1.2445i$   \\
				
				\hline
				$2$&$ -0.3727-1.3909i$ & $ 0.000+0.000i$	\\
				\hline
			     $3$ 	& $0.000+0.000i$ & $-0.3727-1.3909i$ \\
			     	\hline
			     $4$&$ 2.4728+0.6626i$ & $ 0.000+0.000i$	\\
			     \hline
			     $5$ 	& $0.000+0.000i$ & $2.4728+0.6626i$\\
			     	\hline
			     $6$&$ 4.1412+2.206i$ & $ 0.000+0.000i$	\\
			     \hline
			     $7$ 	& $0.000+0.000i$ & $4.1412+2.206i$ \\
			     	\hline
			     $8$&$ -0.6626-2.4728i$ & $ 0.000+0.000i$	\\
			     \hline
			     $9$ 	& $0.000+0.000i$ & $-0.6626-2.4728i$ \\
			     	\hline
			     $10$&$ 1.3909+0.3737i$ & $ 0.000+0.000i$	\\
			     \hline
			     $11$ 	& $0.000+0.000i$ & $1.3909+0.3737i$ \\
			     	\hline
			\end{tabular}
		}
	}
	
\end{table}
For this setting, the channel can be estimated perfectly on both the baseband antenna ports as shown in Table~\ref{tab:Method1_CIR}, thereby allowing for correct data demodulation. 
\begin{table}[H]
	\centering
	{
		\caption{ Effective CIR on \textit{port}-0, \textit{port}-1}
		\label{tab:Method1_CIR}
		\renewcommand{\arraystretch}{1.5}
		{\fontsize{8}{8}\selectfont
			\begin{tabular}{|c|c|c|} \hline 
				\textbf {Time index}	&\textbf {\textit Port0 } &\textbf {\textit Port1 } \\
				
				\hline \hline
				$0$ & $-0.28 $&$-0.28$   \\
				\hline
				$1$ &$ 1$	&$1$   \\
				
				\hline
				$2$&$ -0.28$ & $ -0.28 $	\\
				\hline
			\end{tabular}
		}
	}
	\vspace{10pt}
\end{table}

\end{document}